\documentclass[aip,jcp,reprint,floatfix,superscriptaddress]{revtex4-2}
\usepackage{placeins}
\usepackage{array}[=2016-10-06] 
\usepackage{collcell}
\usepackage{amsmath,amsfonts,amssymb} %
\usepackage{mathtools} %
\ifx\phone\undefined
\else
\let\Phone\phone
\let\phone\relax
\usepackage{wasysym}

\let\phone\Phone
\let\Phone\relax
\fi
\usepackage{bm} %
\usepackage[bbgreekl]{mathbbol} %
\usepackage{graphicx} %
\usepackage[dvipsnames,table]{xcolor} %
\usepackage{multirow,tabularx} %
\usepackage[acronym]{glossaries} %
\usepackage{braket} %
\usepackage[detect-weight=true]{siunitx} %
\usepackage{fancyhdr} %
\usepackage{microtype} %
\usepackage{natmove} 
\usepackage[caption=false]{subfig} %
\usepackage[breaklinks, %
colorlinks, linkcolor=Blue, citecolor=Blue, urlcolor=Blue]{hyperref} %
\usepackage[version=3]{mhchem} %
\usepackage{wrapfig} %
\AtBeginDocument{\usepackage{booktabs}}
\usepackage[linesnumbered,lined,boxed,ruled]{algorithm2e}
\DeclareSIUnit[number-unit-product = {\,}]\cal{cal}
\DeclareSIUnit\erg{erg}
\DeclareSIUnit\au{au}
\def\MyTitle{Open-shell frozen natural orbital approach for quantum eigensolvers}
\def\MyAuthora{Angela F. Harper} %
\def\MyAuthorb{Xiaobing Liu}
\def\MyAuthorc{Scott N. Genin} %
\def\MyAuthord{Ilya G. Ryabinkin} %
\def\MySubject{Electronic structure} %

\hypersetup{%
  pdftitle={\MyTitle{}}, %
  pdfauthor={\MyAuthora{}, \MyAuthorb{}, \MyAuthorc{}, and \MyAuthord}, %
  pdfsubject={\MySubject{}}, %
  pdfkeywords={} %
}

\graphicspath{{pics/}}

\renewcommand{\arraystretch}{1.5}

\newcolumntype{Y}{>{\centering\arraybackslash}X}
\newcolumntype{s}{>{\collectcell\num}X<{\endcollectcell}} 

\newacronym[longplural={degrees of freedom},           %
            firstplural={degrees of freedom (DOF)},    %
            plural={DOF}]{DOF}{DOF}{degree of freedom} %

\newacronym[longplural={equations of motion},           %
            firstplural={equations of motion (EOM)},    %
            plural={EOM}]{EOM}{EOM}{equation of motion} %

\newacronym{TDSE}{TDSE}{time-dependent Schr\"odinger equation}
\newacronym{TIDSE}{TIDSE}{time-independent Schr\"odinger equation}
\newacronym{TDVP}{TDVP}{time-dependent variational principle} %
\newacronym{DFVP}{DFVP}{Dirac--Frenkel variational principle} %
\newacronym{MCTDH}{MCTDH}{multiconfguration time-dependent Hartree}
\newacronym{BCH}{BCH}{Baker--Campbell--Hausdorff}

\newacronym{vMCG}{vMCG}{variational multi-configurational Gaussian} %
\newacronym{FMS}{FMS}{full multiple spawning}                       %
\newacronym{G-MCTDH}{G-MCTDH}{Gaussian-based multi-configuration time-dependent Hartree} %
\newacronym{MAE}{MAE}{mean absolute error} %
\newacronym{pMAER}{\%MAER}{percentage of mean absolute error reduction} %

\newacronym[longplural={random-access memory},            %
            firstplural={random-access memory (RAM)},     %
            plural={RAM}]{RAM}{RAM}{random-access memory} %

\newacronym{GWP}{GWP}{Gaussian wavepacket}           %
\newacronym{TGWP}{TGWP}{thawed Gaussian wavepacket}  %
\newacronym{GHA}{GHA}{global harmonic approximation} %
\newacronym{LHA}{LHA}{local harmonic approximation}  %
\newacronym{ZPE}{ZPE}{zero-point energy}             %
\newacronym{VM}{VM}{vibrational mode}                %
\newacronym{CVM}{CVM}{curvilinear vibrational mode}  %

\newacronym{ML}{ML}{machine learning} %
\newacronym{ML-PES}{ML-PES}{machine-learned potential energy surface} %
\newacronym{KRR}{KRR}{kernel ridge regression}     %
\newacronym{GPR}{GPR}{Gaussian process regression} %

\newacronym{OLED}{OLED}{organic light-emitting diode}     %
\newacronym{NISQ}{NISQ}{noisy intermediate-scale quantum} %
\newacronym{SQD}{SQD}{sample-based quantum diagonalization} %

\newacronym{JW}{JW}{Jordan--Wigner} %
\newacronym{BK}{BK}{Bravyi--Kitaev} %

\newacronym{QPE}{QPE}{quantum phase estimation}          %
\newacronym{VQE}{VQE}{variational quantum eigensolver}   %
\newacronym{QMF}{QMF}{qubit mean-field}                  %
\newacronym{QCC}{QCC}{qubit coupled cluster}             %
\newacronym{iQCC}{iQCC}{iterative qubit coupled cluster} %
\newacronym{iCCSDn}{iCCSDn}{iterative n-body excitation inclusive coupled-cluster single double} %
\newacronym{PQA}{PQA}{parametrized quantum annealing}    %
\newacronym{DIS}{DIS}{direct interaction set}            %
\newacronym[longplural={involutary linear combinations of anti-commuting Paulis}, %
            firstplural={involutory linear combinations of anti-commuting Paulis (ILCAP)}, %
            plural={ILCAP}]{ILCAP}{ILCAP}{involutory linear combination of anti-commuting Paulis} %
\newacronym{DHA}{DHA}{diagonal Hessian approximation} %

\newacronym{CAS}{CAS}{complete active space}    %
\newacronym{PES}{PES}{potential energy surface} %
\newacronym{PEC}{PEC}{potential energy curve}   %
\newacronym{AO}{AO}{atomic orbital}             %
\newacronym{MO}{MO}{molecular orbital}          %
\newacronym{CMO}{CMO}{canonical molecular orbital}   %
\newacronym{FNO}{FNO}{frozen natural orbital}      %
\newacronym{NO}{NO}{natural orbital}                 %
\newacronym{NOON}{NOON}{natural orbital occupation number}  %
\newacronym{OS-FNO}{OS-FNO}{open-shell frozen natural orbital}  %
\newacronym{ZAPT-FNO}{ZAPT-FNO}{ZAPT2 frozen natural orbital}  %

\newacronym{CI}{CI}{configuration interaction}        %
\newacronym{FCI}{FCI}{full configuration interaction} %
\newacronym{CASCI}{CASCI}{complete active space configuration interaction} %
\newacronym{MCSCF}{MCSCF}{multiconfigurational self-consistent field} %
\newacronym{CASSCF}{CASSCF}{complete active space self-consistent field} %
\newacronym{CC}{CC}{coupled cluster}           %
\newacronym{UCC}{UCC}{unitary coupled cluster} %
\newacronym{GUCC}{GUCC}{generalized unitary coupled cluster} %
\newacronym{UCCSD}{UCCSD}{unitary coupled cluster singles and doubles} %
\newacronym{GUCCSD}{GUCCSD}{generalized unitary coupled cluster singles and doubles} %
\newacronym{CCSD}{CCSD}{coupled-cluster singles and doubles} %
\newacronym{CCSD-T}{CCSD(T)}{coupled-cluster singles and doubles and non-iterative triples} %
\newacronym{RHF}{RHF}{restricted Hartree--Fock}                     %
\newacronym{CIS}{CIS}{configuration interaction singles}            %
\newacronym{CISD}{CISD}{configuration interaction singles and doubles} %
\newacronym{ROHF}{ROHF}{restricted open-shell Hartree--Fock}        %
\newacronym{UHF}{UHF}{unrestricted Hartree--Fock}                   %
\newacronym{MPS}{MPS}{matrix product states}                        %
\newacronym{DMRG}{DMRG}{density-matrix renormalization group}       %
\newacronym{DFT}{DFT}{density-functional theory}                    %
\newacronym{TDDFT}{TDDFT}{time-dependent density-functional theory} %
\newacronym{ENPT}{ENPT}{Epstein-Nesbet perturbation theory}         %
\newacronym{MP}{MP}{M{\o}ller--Plesset perturbation theory}         %
\newacronym{MP2}{MP2}{second-order M{\o}ller--Plesset perturbation theory} %
\newacronym{ZAPT}{ZAPT}{Z-averaged perturbation theory} %
\newacronym{ZAPT2}{ZAPT2}{second-order Z-averaged perturbation theory} %
\newacronym{MRMP2}{MRMP2}{second-order multi-reference M{\o}ller--Plesset perturbation theory} %

\newacronym{SQP}{SQP}{sequential quadratic programming} %
\newacronym{MMA}{MMA}{method of moving asymptotes}      %

\def\be{\begin{equation}} %
\def\ee{\end{equation}} %
\def\bea{\begin{eqnarray}} %
\def\eea{\end{eqnarray}} %

\begin{document}

\title{\MyTitle}

\author{\MyAuthora}
\thanks{\href{mailto:angela.harper@otilumionics.com}{angela.harper@otilumionics.com} \\ \textsuperscript{\dag}Deceased: 11 December 2025}

\author{\MyAuthorb}
\author{\MyAuthorc}

\author{\MyAuthord\textsuperscript{\dag}}%
\affiliation{OTI Lumionics Inc., 3415 American Drive Unit~1, \\
Mississauga, Ontario L4V\,1T4, Canada}

\date{\today} 

\ifx\latin\undefined %
\newcommand{\latin}{\textit} %
\fi %

\begin{abstract}
  We present an open-shell \gls{FNO} approach, which utilizes the \gls{ZAPT2}, to reduce the \acrlong{ROHF} virtual space size with controllable accuracy.
  Our \gls{ZAPT-FNO} selection scheme significantly outperforms the \acrlong{CMO} virtual space truncation scheme based on Hartree--Fock orbital energies, especially when using large multiple-polarized and augmented basis sets.
  We demonstrate that the \Gls{ZAPT-FNO}-selected virtual orbitals lead to a systematic convergence of the correlation energies, but more importantly to the singlet-triplet T$_1$-S$_ 0$ energy gaps with respect to the \gls{CAS} [occupied + virtual] size.
  We confirm our findings by simulating T$_1$-S$_ 0$ gaps in \ce{H2O2} and \ce{O2} molecules using the traditional \gls{CASCI} approach, as well as in stretched \ce{CH2}, for which we also employed the \gls{iQCC} method as a quantum eigensolver.
  Finally, we applied the \gls{iQCC} method with \gls{ZAPT-FNO}-selected active space to the phosphorescent \ce{Ir(ppy)3} complex with \num{260} electrons, where extended basis sets are required to achieve chemical (\latin{ca.~\SI{1}{\milli\hartree}}) accuracy.
  In this case, \gls{CASCI} results are not available; however, the \gls{iQCC}-computed T$_1$-S$_0$ gaps show robust convergence with enlarging basis set and \gls{CAS} size, approaching the experimental value.
  Thus, the \gls{ZAPT-FNO} method is very promising for improving the accuracy of quantum chemical modelling in a resource-efficient manner, and opens the door to simulating open-shell states of large materials within realistic active space sizes and without compromising on basis-set quality.
\end{abstract}

\maketitle

\glsresetall 

\section{Introduction}
\label{sec:introduction}
Open-shell systems are ubiquitous in practical applications of transition metal complexes.
For example, phosphorescent emitters, vital to the multi-billion dollar \gls{OLED} industry, often have ground or low-lying electronic states with unpaired electrons.
Accurate \latin{ab initio} modelling of their electronic structure requires large (extended) \emph{atomic} basis sets with multiple polarization shells and diffuse orbitals.
This results in a huge number of \emph{molecular} orbitals, prompting researchers to look for methods with lower computational costs.
One such method is the \gls{FNO} approach~\cite{kumar2017frozen,taube2005frozen,manishafrozen,taube2008frozen,landau2010frozen,verma2021scaling}.
Unfortunately, existing implementations are largely limited to closed-shell systems; the only exception to this being Ref.~\citenum{pokhilko2020extension}; however their use of an \acrlong{UHF} reference, makes this approach incompatible with most common quantum eigensolvers. 

Solving difficult electronic structure problems in quantum chemistry is a powerful driving force for the development of quantum algorithms and hardware; ultimately the aim is to achieve a quantum advantage over classical methods (see, however, a critical opinion in Ref.~\citenum{Lee:2023/natcommun/1952}).
Among the various approaches proposed for \gls{NISQ} devices~\cite{kandala2017hardware}, the \gls{VQE}~\cite{peruzzo2014variational,wecker2015progress} has emerged as a leading framework, due to its adaptability to hardware limitations and compatibility with fermionic Hamiltonians.
The \gls{iQCC} method~\cite{ryabinkin2018qubit,ryabinkin2020iterative,ryabinkin2021posteriori,ryabinkin2023efficient,ryabinkin2025optimization} is a variant of \gls{VQE} for variational energy minimization of qubit Hamiltonians on classical hardware.
This ``quantum-on-classical'' approach achieves linear scaling with the size of the qubit Hamiltonian, making it a practical tool for simulating large quantum systems.

In addition to emulating quantum-hardware calculations, \gls{iQCC} has demonstrated competitive accuracy with classical quantum chemistry methods such as \gls{DFT} and \gls{CCSD} across a variety of chemical systems~\cite{genin2022estimating,ryabinkin2023efficient}.
A key advantage of \gls{iQCC} is its ability to simulate Hamiltonians of sizes that exceed the capabilities of current quantum hardware.
For context, the largest simulated quantum computation to-date is the Hartree--Fock simulation of the \ce{H12} hydrogen chain from Google AI Quantum, using \num{12} logical qubits in a CAS(12,12) active space~\cite{google2020hartree}.
By comparison, \gls{iQCC} has simulated systems of up to \num{72} qubits [or CAS(36,36)~\cite{genin2022estimating}], and has the potential to simulate even larger systems \cite{genin2025quantumadvantagechemistry}, limited only by the capabilities of modern CPU hardware~\cite{ryabinkin2025optimization}.
Nevertheless, in both the Google AI Quantum and \gls{iQCC} case, the size of the initial qubit Hamiltonian was reduced from including all electrons and orbitals, in order for the computation to be feasible within the limits of the quantum or classical hardware on which it was simulated.
This is a common practice in quantum chemistry, called ``active space selection'', in which the total number of orbitals is reduced to a smaller set which contribute the most to the total energy of the system.
The goal of active space methods is to select a set of orbitals which are both chemically relevant, and also computationally tractable, such that the final Hamiltonian is small enough to be simulated on the available hardware.

Active space selection is an established field in quantum chemistry~\cite{stein2019autocas,lei2021icas,kaufold2023automated}, as performing many body calculations which scale as $\mathcal{O}$(N$^7$), is only feasible for small systems and basis sets.
While occupied orbitals are well-defined across basis set representations, virtual orbitals can form a dense manifold of states above the highest occupied molecular orbital~\cite{schmidt2015valence}.
In augmented basis sets, the virtual space contains many diffuse orbitals which form a continuum of orbital states that are close in energy, making it difficult to extract chemically relevant low-lying virtual states.
Therefore, several methods aim at reducing the size of the valence-virtual space, to a selected active set of orbitals, while simultaneously retaining the most chemically relevant orbitals.

 For example, the optimized orbital \gls{CCSD} method yields so-called Br\"{u}ckner orbitals~\cite{sherrill1998energies,brueckner1954nuclear}, in which the correlation energy is minimized with respect to orbital variations between occupied and virtual orbitals in the active space.
            Additional methods for virtual space compression {rely on local correlation techniques, includ{ing} the domain-based local pair natural orbital (DLPNO) approach \cite{neese2009efficient,neese2009efficient1,riplinger2013efficient}, the local natural orbital (LNO) scheme \cite{rolik2013efficient}, and methods utilizing orbital-specific virtuals (OSVs) \cite{yang2012orbital} or projected atomic orbitals (PAOs) \cite{pulay1983localizability,schutz2001low}.
            {These local methods reduce the scaling of post-Hartree-Fock calculations, enabling correlation energy recovery in systems containing hundreds or even thousands of atoms} with computational efficiency.
            {Another approach to virtual space compression is t}he \gls{FNO} approach~\cite{taube2005frozen,taube2008frozen}{, which computes the \glspl{FNO} as the eigenvectors of the virtual-virtual block of the total \gls{MP2} {one-particle }density matrix~\cite{landau2010frozen}.
            {By contrast, pair natural orbital approaches are a special case of FNO, where the virtual density matrix is constructed and diagonalized for specific occupied orbital pairs~\cite{lowdin1955quantum,neese2009efficient1,neese2009efficient}.}
            Finally, the AB2 (antibonding second-order perturbation) method has been proposed as an approach to constructing valence antibonding orbitals by maximizing opposite-spin \gls{MP2} pair correlation amplitudes, yielding a robust way to define molecule-adapted minimal bases~\cite{aldossary2022non}.

As an alternative to perturbation theory approaches, there are several strategies which calculate orbital entanglement to identify strongly correlated orbitals \cite{legeza2003controlling,boguslawski2013orbital}.
These methods employ \gls{DMRG} in a large initial orbital space to compute natural orbitals and their entanglement signatures.
This enables systematic selection of the relevant active space based on occupation numbers or entanglement entropy, as implemented in the AutoCAS method \cite{stein2016automated,stein2019autocas}, which is particularly useful for multi-reference calculations.

Of the aforementioned approaches for truncating the virtual space, the \gls{FNO} approach is well-suited for quantum computing applications, as it has been extensively used for closed-shell restricted Hartree Fock orbital selection \cite{taube2005frozen,taube2008frozen}.
The resulting set of orbitals from such a restricted method can easily be transformed by standard fermion to qubit transformations~\cite{nielsen2005fermionic}, to be suitable for input to a quantum eigensolver, such as \gls{iQCC}.
To extend the \gls{FNO} approach to open-shell systems, we propose a variant based on \gls{ZAPT2}~\cite{fletcher2002gradient,aikens2006scalable}, which uses a \gls{ROHF} reference.
This simple but effective extension to the \gls{FNO} approach, opens up avenues for computing energy gaps in molecules with Rydberg orbitals and strong correlation, where diffuse basis functions are essential to recover accurate ground state energies.

The manuscript is organized as follows.
We first present an overview of the \gls{FNO} method and \gls{ZAPT2} perturbation theory, and describe how their combination yields a practical \gls{ZAPT-FNO} approach compatible with quantum computing methods.
We then benchmark \gls{ZAPT-FNO} against \gls{CMO} selection for several systems, using \gls{iQCC} to perform the energy minimization.
For \ce{H2O2}, we demonstrate that \gls{ZAPT-FNO} recovers correlation energy more efficiently in a smaller active space size, particularly for augmented basis sets.
Additionally, we demonstrate that the \ce{O2} triplet-singlet (T$_1$-S$_0$) energy gap converges smoothly with respect to \gls{CAS} size for the \gls{ZAPT-FNO} approach, while the \gls{CMO} energy gaps \textit{may} achieve convergence only as a result of fortuitous error cancellation.
We further validate our approach by computing the singlet and triplet states of stretched \ce{CH2} across a range of \ce{C-H} distances, and achieve chemical accuracy relative to \gls{CASCI} calculations with a higher quality basis set than previously reported~\cite{liepuoniute2025quantum} in the same active space size.
Finally, we consider the \num{260} electron (61 atoms) \ce{Ir(ppy)_3} complex, which is a challenging system to simulate due to both its size, and the Jahn-Teller distortion in the triplet state~\cite{fine2012electronic}.
Despite the complexity of the system, \gls{ZAPT-FNO} enables accurate prediction of the T$_1$–S$_0$ gap using a compact active space, thereby illustrating the method’s promise for simulating large complexes with chemical accuracy.

\section{Methods}
\label{sec:methods}

\subsection{\Acrfull{ZAPT2}}
\label{sec:ZAPT2-methods}

The \gls{ZAPT2} correlation correction to the energy is an extension of the \gls{MP2} perturbation theory to open-shell systems~\cite{fletcher2002gradient,aikens2006scalable}.
The correction is based on the closed-shell \gls{MP2} theory, with differences arising from the necessity for treatment of the singly occupied orbitals.
The \gls{ZAPT2} energy correction is expressed in the \gls{MO} basis, as a sum over two-electron integrals $(pq|rs)$ and their amplitudes.
In the case of a closed shell system, the $E^{(2)}_{\text{ZAPT}}$ energy correction is equivalent to the $E^{(2)}_{\text{MP}}$ energy correction, as we will show below.

In the following equations, orbital indices are defined as follows: $i$,$j$,$k$ index doubly occupied active \glspl{MO}, $w$,$x$,$y$,$z$ index singly occupied active \glspl{MO}, and $a$,$b$,$c$ index virtual, unoccupied active \glspl{MO}.
General \glspl{MO} are indexed by $p$,$q$,$r$,$s$, and \glspl{AO} are indexed by $\mu$,$\nu$,$\lambda$,$\sigma$.
The \gls{ZAPT2} energy correction is then given by the following expression,
\begin{align}
E_{\text{ZAPT}}^{(2)} = &\ \frac{1}{2} \sum_{i,j} \sum_{p,q}^{\mathit{s.v.}}
\frac{(ip|jq)[C_{pq}(ip|jq) - (iq|jp)]}{D_{ij}^{pq}} \notag \\
&+ \frac{1}{2} \sum_{p,q}^{\mathit{d.s.}} \sum_{a,b}
\frac{(pa|qb)[C_{pq}(pa|qb) - (pb|qa)]}{D_{pq}^{ab}} \notag \\
&+ \sum_{ixya} (ix|ya)\frac{(ix|ya)}{D_{iy}^{xa}}
+ \sum_{ixya} (ix|xa)\frac{(iy|xa)}{D_{ii}^{aa}}
\label{eq:zapt2-energy}
\end{align}
where for the set of singly occupied orbitals (SOCC) ,
\begin{equation}
C_{pq} =
\begin{cases}
1, & \text{for both } p \in \text{SOCC} \text{ and } q \in \text{SOCC}, \\
2, & \text{otherwise},
\end{cases}
\end{equation}
and denominators are a summation over the orbital energies, $\varepsilon$, defined as
\begin{equation}
  D_{pq}^{rs} = \varepsilon_p + \varepsilon_q - \varepsilon_r' - \varepsilon_s',
\end{equation}
The notation \textit{s.v.}
and \textit{d.s.}
refers to summations over the singly occupied and virtual indices, and doubly occupied and singly occupied indices, respectively.
The orbital energies for the singly occupied orbitals include an additional integral component, which is a sum of exchange integrals over the singly occupied orbitals:
\begin{align}
  \varepsilon_p &= \varepsilon_{pp} - \frac{1}{2} \sum_y (py | py ), && \text{for } p \in SOCC, \notag \\
  \varepsilon'_p &= \varepsilon_{pp} + \frac{1}{2} \sum_y (py | py), && \text{for } p \in SOCC, \notag \\
  \varepsilon'_p &= \varepsilon_p = \varepsilon_{pp}, && \text{for } p \notin SOCC,
\end{align}
where $\varepsilon_{pp}$ is the orbital energy of the $p$th orbital in the \gls{MO} basis.
The third and fourth terms in Eq.~\ref{eq:zapt2-energy} are summations over the singly occupied orbitals, and are only present in the open-shell case.
If we consider the closed-shell case (\latin{i.e.}
no singly-occupied orbital indices, $\varepsilon_p' = \varepsilon_p$, and $C_{pq}=2$ ), we can see that Eq.~\ref{eq:zapt2-energy} reduces to the \gls{MP2} energy correction~\cite{shavitt2009many}, $E^{(2)}_{\text{MP}}$,
\begin{align}
  E_{\text{MP}}^{(2)} = &\ \sum_{a,b} \sum_{i,j} \frac{(ij|ab)[2(ij|ab)-(ij|ba)]}{\varepsilon_i + \varepsilon_j - \varepsilon_a - \varepsilon_b}. \notag \\
  \label{eq:mp2-energy}
\end{align}
Given the above similarities between \gls{ZAPT2} and \gls{MP2} derivations, there is a natural extension of the \gls{FNO} approach, described in the following section, to open-shell systems using \gls{ZAPT2}.

\subsection{\Acrfull{FNO} approach}
\label{sec:fno-approach}

The \gls{FNO} approach is a well-established method for truncating the virtual orbital active space in correlated wavefunction methods such as \gls{CCSD} and \gls{CCSD-T}, and has been widely adopted for closed-shell systems~\cite{taube2005frozen,kumar2017frozen,deprince2013accurate}.
However, direct extension of the \gls{FNO} approach to open-shell systems is not straightforward.
Pokhilko and co-authors~\cite{pokhilko2020extension} showed that using an unrestricted Hartree--Fock reference leads to inconsistent truncation of the $\alpha$ and $\beta$ virtual orbital spaces, which causes ``erratic'' behavior in the final total energies.
They solve this problem by introducing a new extension to the \gls{FNO} approach using the equation-of-motion coupled cluster approach, which uses a singular value decomposition of the $\alpha$ and $\beta$ orbitals overlap matricies to treat the singly occupied orbital space, and remove the inconsistencies between the $\alpha$ and $\beta$ orbital channels.
The equation-of-motion \gls{FNO} method, implemented in Q-Chem~\cite{epifanovsky2021software}, has been shown to yield accurate results for the triplet-singlet energy gaps in small molecules, such as ethene, methanol, and cyclobutadiene, with errors in the gap energy of less than \SI{2}{\milli\hartree}~\cite{manishafrozen}.
While effective for the \gls{CCSD-T} application case, the equation-of-motion \gls{FNO} approach suffers from the fact that the resulting unrestricted reference contains two spin channels, and is therefore incompatible with fermion-to-qubit mappings for single fermionic Hamiltonians.

To address this limitation, we propose an alternative \gls{ZAPT-FNO} scheme based on a \gls{ROHF} reference, using \gls{ZAPT2} perturbation theory~\cite{fletcher2002gradient,aikens2006scalable}.
As a spin-restricted method, \gls{ZAPT2} yields a single set of orbitals and is thus naturally compatible both with the original \gls{FNO} formalism and with standard fermion-to-qubit mappings.
Below, we outline the general \gls{FNO} algorithm which is applicable both to \gls{RHF} and \gls{ROHF} references, using either \gls{MP2} or \gls{ZAPT2} to compute the \glspl{NO}.
The algorithm is as follows:

\begin{enumerate}
\item Perform a self-consistent field calculation using either an \gls{RHF} or \gls{ROHF} reference to obtain the \glspl{CMO}
\item Compute the virtual-virtual part of the \gls{MP2} or \gls{ZAPT2} one-particle density matrix ($P^{(2)}$)
\item Diagonalize the virtual-virtual part of $P^{(2)}$ to obtain the \glspl{NO}, sorted in decreasing order of occupation
\item Select the desired number of \glspl{NO} to retain, either using a threshold of occupation or a set number of orbitals.
  The remaining virtual orbitals with occupation number lower than this threshold are frozen (\latin{i.e.}
  not used).
\item Semicanonicalize the active (unfrozen) virtual orbitals by diagonalizing the active-active block of the Fock operator constructed from the \gls{ROHF} or \gls{RHF} reference. This transformation preserves the spin-restriction of the spatial orbitals.
\item Transform the one- and two-electron integrals into the new
  orbital basis of semicanonicalized orbitals for all active orbitals.
\item Perform a transformation from the fermion to qubit
  representation of the final active space, and perform the \gls{iQCC}
  calculation in the new basis.
\end{enumerate}

Details of how to compute the \gls{ZAPT2} one-particle density matrix are given in Appendix ~\ref{sec:zapt2-one-particle-density-matrix}.
As a result of truncating the active space \glspl{NO} in Step~4, the final energy calculated by \gls{iQCC} does not include the contributions from the frozen orbitals.
However, because the \glspl{NO} are selected based on their contributions to the correlation energy, a correction can be applied to the final energy to recover the energy of the frozen orbitals.
The correction is computed as the difference between the \gls{MP2} (or \gls{ZAPT2}) correlation energy (defined in Eqs.~\ref{eq:zapt2-energy} and \ref{eq:mp2-energy}) evaluated in the full \gls{MO} basis and the truncated \gls{NO} basis,
\begin{align}
\Delta E_{\text{FNO}} &= E_{\text{MP/ZAPT}}^{\text{MO}} - E_{\text{MP/ZAPT}}^{\text{NO}} \label{eq:energy-correction} \notag \\
\end{align}
such that the corrected \gls{iQCC} energy is given by,
\begin{align}
E_{\mathrm{\gls{iQCC}}}^{\text{FNO}} &= E_{\mathrm{\gls{iQCC}}}^{\text{NO}} + \Delta E_{\text{FNO}}. \notag \\
\end{align}

This correction allows systematic improvement of the \gls{iQCC} minimized energy by accounting for dynamical correlation outside the active space.
Implementing the \gls{ZAPT-FNO} scheme for the open-shell \gls{ZAPT2} case thus requires that we compute two main quantities: the virtual-virtual block of the \gls{ZAPT2} one-particle density matrix (P$^{(2)}$), and the corresponding \gls{ZAPT2} correlation energy for both the selected \glspl{NO} and the full set of \glspl{MO}.

We note that due to the \gls{ZAPT2} construction, and semicanonicalization in the virtual space, that outside of iQCC, this approach will be applicable primarily to iterative procedures such as CCSD. 
One should be cautious when applying the ZAPT-FNO scheme to perturbative corrections such as the (T) correction in CCSD(T), as the single set of semicanonicalized virtual orbitals will not converge to the canonical CCSD(T) energy \cite{watts1993coupled,scuseria1991open}. 

\subsection{Outline of \gls{iQCC}}
\label{sec:iQCC-methods}

To avoid self-repetition, we refer the reader to our recent work outlining the \gls{iQCC} approach in detail at Ref. \citenum{ryabinkin2025optimization}, and provide only a brief overview in this section. 

The \gls{iQCC} approach targets the ground-state energy of an active-space second-quantized Hamiltonian~\cite{peruzzo2014variational,ryabinkin2023efficient},
\begin{equation}
\label{eq:hamiltonian}
\hat{H}_e = E_{\text{core}} + \sum_{pq} f_{pq} \hat{a}^\dagger_p \hat{a}_q + \tfrac{1}{2} \sum_{pqrs} g_{pqrs} \hat{a}^\dagger_p \hat{a}^\dagger_q \hat{a}_r \hat{a}_s,
\end{equation}
where $E_{\text{core}}$ is the energy of the frozen-core orbitals, $\hat{a}^\dagger_p$ and $\hat{a}_q$ are the creation and annihilation operators, and $f_{pq}$, $g_{pqrs}$ are one- and two-electron integrals over the chosen active space.

Using a \gls{JW} transformation~\cite{nielsen2005fermionic}, the Hamiltonian is mapped to,
\begin{equation}
\hat{H} = \sum_{k=1}^M C_k \hat{P}_k,
\end{equation}
where $\hat{P}_k$ are Pauli words and coefficients $C_k$ are determined by the electronic integrals.

Finally, one must choose the initial reference vector $\ket{0}$, which is a direct product state of spin-orbitals,
\begin{equation}
  \label{eq:reference-state}
  \ket{0} = \prod_{k = 1}^{n_e} \ket{\downarrow}_k \prod_{k=1}^{n-n_e} \ket{\uparrow}_k,
\end{equation}
where $n_e$ is the number of electrons in the active space, and $n$ is the total number of spin-orbitals in the active space, which determines the total number of qubits in the system.

The \gls{iQCC} algorithm then aims to variationally minimize the energy of the Hamiltonian, $\hat{H}$, via
\begin{equation}
\label{eq:iQCC-minimization}
E_{\mathrm{iQCC}}[\hat{T}] = \min_{t} \bra{0} \hat{U}^\dagger(t) \hat{H} \hat{U}(t) \ket{0},
\end{equation}
where $\hat{U}(t)$ is a unitary operator realizable on a quantum computer.
The \gls{iQCC} approach iteratively optimizes the amplitudes $t$ of the unitary operator, via an iterative process of generator selection, energy minimization, and amplitude updating.
Further details of this method, are provided in Refs.~\citenum{ryabinkin2018qubit,ryabinkin2020iterative,ryabinkin2021posteriori,ryabinkin2023efficient}, and improvements to the amplitude optimization as well as a detailed overview of the iQCC approach are provided in Ref.~\citenum{ryabinkin2025optimization}.

It is the reference state, given in Eq.~\ref{eq:reference-state}, that defines the \gls{CAS} size and therefore the number of qubits, and ultimately computational cost of the calculation.
The \gls{CAS} is usually defined as CAS(n$_e$,n$_o$), where n$_e$ is the number of electrons in the active space, and n$_o$ is the number of spin-orbitals in the active space.
The number of qubits required for this active space is $n = 2n_o$, as each spin-orbital is represented by a single qubit in the \gls{JW} transformation.
By improving the quality of the fermionic Hamiltonian given in Eq.~\ref{eq:hamiltonian}, as well as limiting the \gls{CAS} size by obtaining a higher quality set of orbitals, we aim to improve the convergence of the \gls{iQCC} energy minimization, and ultimately the accuracy of the final energy.

\section{Results}
\label{sec:results}

\subsection{Computational Details}
\label{sec:comp-deta}

The implementation of computing the \gls{ZAPT2} energy correction and the one-particle density matrix in the \gls{MO} basis was done using the \texttt{PySCF}~\cite{sun2020recent} package, which includes the ability to access the integrals in the \gls{MO} basis, and perform the necessary transformations from the \gls{AO} to \gls{MO} basis.
The \gls{ZAPT-FNO} approach was implemented in \texttt{python}~v3.13, using the \texttt{numpy} library for efficient handling of large arrays of the one and two-electron integrals, and parallelized with \texttt{mpi4py}~\cite{dalcin2021mpi4py}. 
While the primary focus of our implementation is accuracy, and reduced orbital requirements, especially in the open shell triplet state, we provide details of the computational cost of our workflow in Appendix~\ref{sec:cpu-scaling}.

We compare results from the \gls{ZAPT-FNO} approach to those obtained using \glspl{CMO}, where \glspl{CMO} refer to Hartree–Fock optimized orbitals without any subsequent orbital optimization (i.e. no \gls{CASSCF} calculations were performed). 
When \gls{CASCI} is used as a comparison eigensolver, it represents a CI calculation within the defined \gls{CAS} (constructed from either \glspl{CMO} or \gls{ZAPT-FNO} orbitals as described in the text).

All \gls{iQCC} calculations were performed using a C++ implementation of \gls{iQCC}~\cite{ryabinkin2020iterative,genin2025quantumadvantagechemistry} which enabled parallelization of the energy minimization, and was run on a single node with \num{128}~cores, using a maximum of \SI{1.02}{\tebi\byte} of \gls{RAM} during the energy minimization in the case of the largest system [CAS(40,40) \ce{Ir(ppy)_3}].
Details of the specific C++ implementation and computational requirements are available in full from \citet{genin2025quantumadvantagechemistry}.
The one- and two-electron integrals generated using either the \gls{CMO} or \gls{FNO} approach are used to construct the qubit Hamiltonian using the \gls{JW} transformation~\cite{nielsen2005fermionic}.
For each system, a fixed number of \gls{iQCC} steps were performed starting from the qubit Hamiltonian such that the optimized amplitudes of the qubit coupled cluster Ansatz are below \num{0.015}. 
This convergence protocol is consistent with previous studies using \gls{iQCC} \cite{genin2022estimating}, and provides a way to fairly compare the results from multiple \gls{iQCC} optimizations.

\subsection{\ce{H2O2} case study: recovering correlation with \gls{ZAPT-FNO}}
\label{sec:ceh2o2-case-study}

In general, freezing out the highest energy canonical Hartree--Fock \glspl{MO} in a system, leads to large errors in the electron-correlation energies, $E^{(2)}$. \citet{kumar2017frozen} show that using the \gls{FNO} approach for the closed shell system of hydrogen peroxide (\ce{H2O2}) with the aug-cc-pVTZ~\cite{dunning1989gaussian,kendall1992a}, results in a significantly lower error in the final energy, which can be further decreased with the use of the energy correction given by Eq.~\ref{eq:energy-correction}.

An equivalent test is shown in Fig.~\ref{fig:h2o2-zapt-correction}, for the high spin triplet state (T$_1$) of \ce{H2O2}, using the \gls{ZAPT-FNO} approach. The \ce{H2O2} geometry used in this calculation was extracted from Ref.~\citenum{kumar2017frozen}, in which the structure was geometry optimized with the B3LYP functional~\cite{becke1993density}, with an aug-cc-pVDZ basis set~\cite{dunning1989gaussian}. We compare the calculated $E_{\text{ZAPT}}^{(2)}$ for increasing number of frozen virtual orbitals, both for the \gls{ZAPT-FNO} approach, and for the \gls{CMO} approach and observe that for an equivalent number of frozen virtuals, we recover a much larger fraction of the correlation energy with the \gls{ZAPT-FNO} approach than for the \gls{CMO} approach.
This follows the same trend as the closed-shell \ce{H2O2} case~\cite{kumar2017frozen}, and demonstrates that using the \gls{ZAPT2} theory to select the open-shell virtual orbitals is a viable option for open-shell systems, as it recovers correlation energy similarly to the closed-shell \gls{RHF} reference case.

\begin{figure}
  \centering 
  \includegraphics[width=0.50\textwidth]{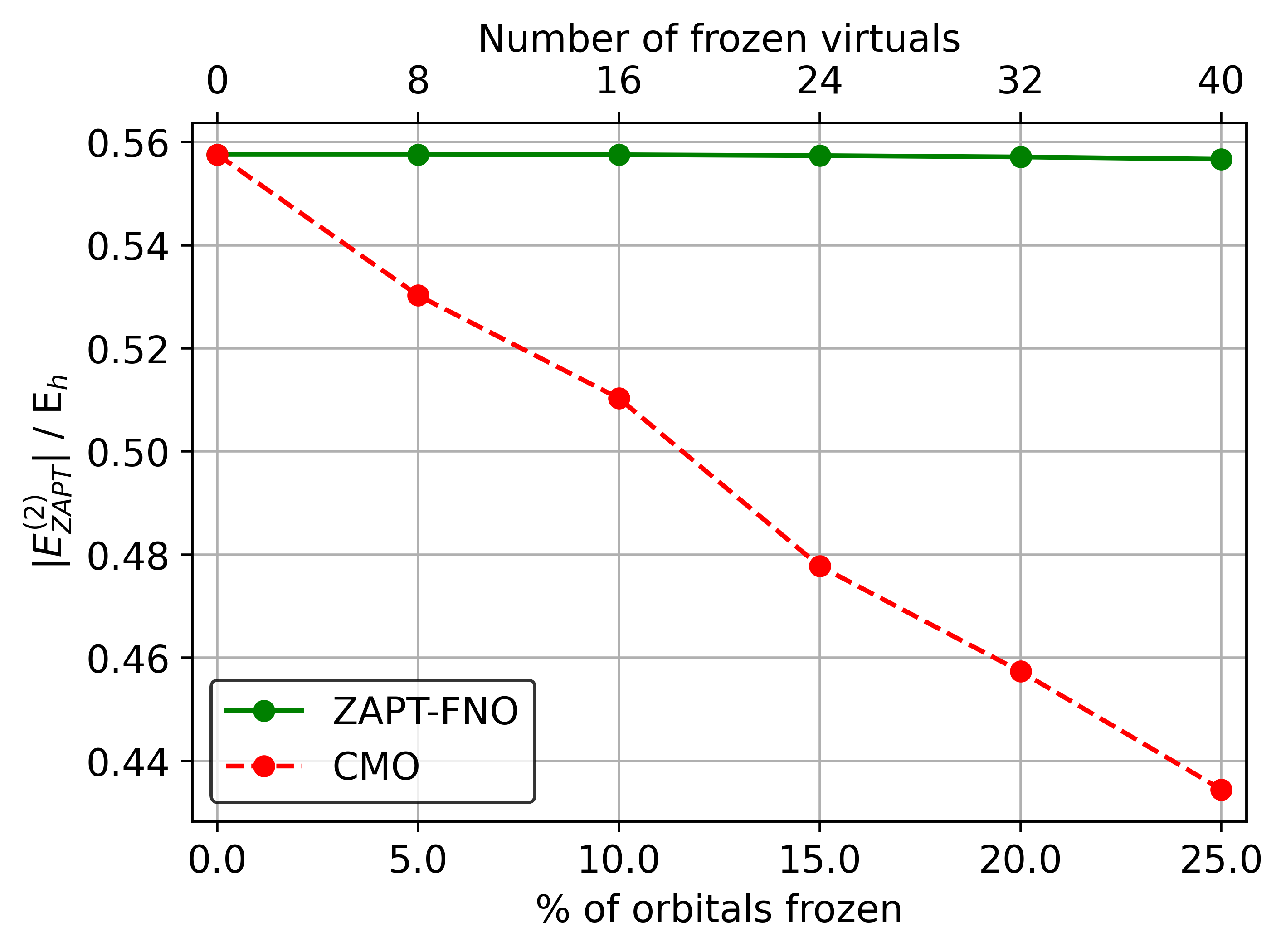}
  \caption{The total correlation energy, $E_{\text{ZAPT}}^{(2)}$ computed at an increasing number of frozen virtuals for both the \gls{CMO} and \gls{ZAPT-FNO} approaches using the aug-cc-pVTZ basis set on the high spin triplet state of \ce{H2O2}.
  The total correlation energy at an increasing number of frozen virtuals decreases rapidly for the \gls{CMO} orbitals, while a large amount of the correlation energy is recovered using the \gls{ZAPT-FNO} approach.}
  \label{fig:h2o2-zapt-correction}
\end{figure}

By recovering a larger amount of the correlation energy in a smaller number of active virtual orbitals for the open-shell systems, the T$_1$-S$_0$ energy gap converges using a smaller fraction of the virtual space.
In Fig.~\ref{fig:h2o2-gap} we compute the T$_1$-S$_0$ energy gap at the \gls{ZAPT2} level of theory for \ce{H2O2} with the aug-cc-pVTZ basis set~\cite{dunning1989gaussian,kendall1992a}, using an increasing number of \glspl{MO} up to the full set of \num{160} \glspl{MO} (\num{0} frozen virtuals).
The energy gap is computed as
\begin{align}
E^{\text{gap}} &= E^{\text{T}_1} - E^{\text{S}_0} \notag \\
\end{align}
where
\begin{align}
  E^{\text{T}_1} &= E^{\text{T}_1}_{\text{ROHF}} + E^{\text{T}_1 (2)}_{\text{ZAPT}}, \notag \\
  E^{\text{S}_0} &= E^{\text{S}_0}_{\text{ROHF}} + E^{\text{S}_0 (2)}_{\text{ZAPT}}.
\end{align}
By freezing up to \SI{25}{\percent} of the active space with the \gls{FNO} approach, we achieve a T$_1$-S$_0$ gap within \SI{1}{\milli\hartree} of the $E^{\text{gap}}_{\text{ZAPT}}$ value using all \num{160} orbitals in the basis set.
The \gls{CMO} with high-energy frozen orbitals does not converge until 95\% of the \glspl{MO} are used.
Simply truncating the active space by removing the highest energy orbitals cuts out many orbitals which actually have \gls{NO} occupation, and play a large role in the total correlation energy, and subsequently the T$_1$-S$_0$ gap.
\begin{figure} 
  \centering \includegraphics[width=0.48\textwidth]{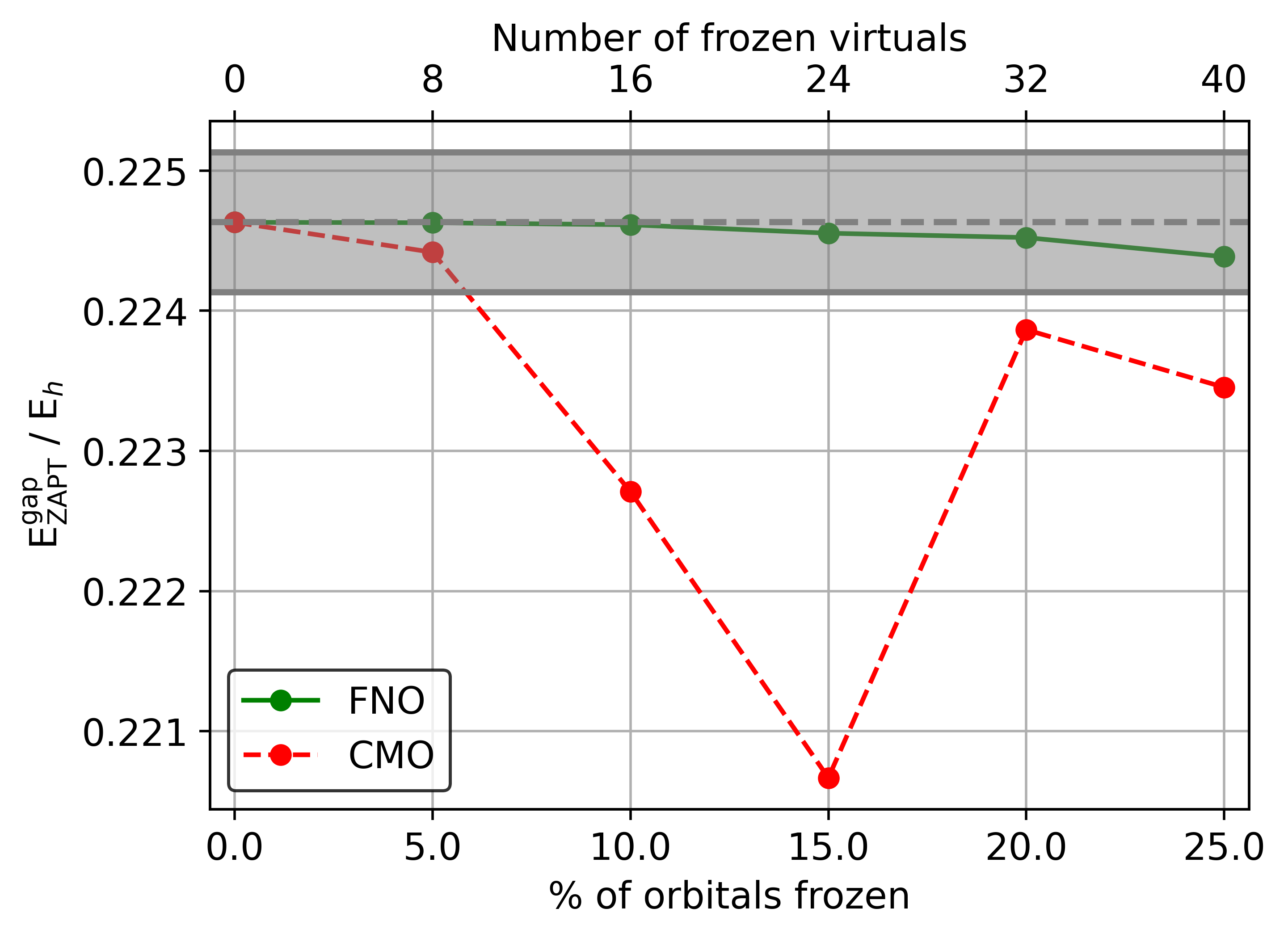}
  \caption{The T$_1$-S$_0$ energy gap for \ce{H2O2} with the aug-cc-pVTZ basis set using the \gls{FNO} approach and the \gls{CMO} approach.
    The grey box indicates ``chemical accuracy'' of \SI{1}{\milli\hartree}.
    The \gls{FNO} $E^{\text{gap}}$ is within \SI{1}{\milli\hartree} at \SI{20}{\percent} frozen virtuals, while the \gls{CMO} approach does not converge to within \SI{1}{\milli\hartree} until the full set of MOs is used.}
  \label{fig:h2o2-gap}
\end{figure}

\subsection{\ce{O2} case study: \gls{CAS} size selection}
\label{sec:O2-case-study}

Active space size selection underpins the accuracy of both total energies and the T$_1$-S$_0$ energy gap.
In this section, we consider how the energy gap converges as a function of the active space size for \ce{O2} with the aug-cc-pVTZ basis set, both for \gls{CMO} and \gls{ZAPT-FNO} selected orbitals.
We use \ce{O2} as a benchmark system to validate the internal consistency and systematic convergence of our \gls{ZAPT-FNO} approach against reference computational methods. 
While the S$_0$ state of \ce{O2} has significant multireference character that cannot be accurately described by single-reference methods with respect to experiment,
we use CCSD(T) here solely as a single‑reference baseline, because it is the highest‑accuracy single-reference method available and provides a consistent and widely accepted reference point~\cite{amsler2023classical}.
This comparison demonstrates 
that \gls{ZAPT-FNO} maintains consistency with standard quantum chemistry methods for challenging molecular systems.

The \ce{O2} molecule was constructed with an \ce{O-O} distance of \SI{1.2075}{\angstrom}, based on the ground state geometry reported in the Computational Chemistry Comparison and Benchmark Database (CCCBDB)~\cite{huber2013molecular,nist2022reference}.
All \ce{O2} qubit Hamiltonians were optimized with iQCC using \num{10} QCC iterations with \num{10} entanglers at each iteration at a maximum order of \num{8}, and a final step with \num{200000} entanglers at a maximum order of \num{1}.

Figure~\ref{fig:o2-gap-convergence}a shows the \glspl{NOON} for the virtual orbitals of \ce{O2} with the aug-cc-pVTZ basis set~\cite{dunning1989gaussian,kendall1992a}, for the S$_0$ and T$_0$ states.
The orbitals used for \gls{CAS} sizes of (8,8), (8,16), (8,22), and (8,30) are shown by horizontal dashed lines, indicating that all orbitals to the left of that line are included in the given active space using the \gls{ZAPT-FNO} approach.
The stepped structure of the \glspl{NOON} indicates natural \gls{CAS} sizes, based on the decrease in occupation number as a function of the orbital index.
The orbitals in the \gls{CMO} approach on the other hand are selected based on their orbital energy rather than their occupation number.
Figure~\ref{fig:o2-gap-convergence}b shows the convergence of the T$_1$-S$_0$ energy gap as a function of the \gls{CAS} size using both the \gls{CMO} selected orbitals and the \gls{ZAPT-FNO} selected orbitals both with and without the $\Delta E_{\text{FNO}}$ correction.
We find that the \gls{ZAPT-FNO} approach has smooth convergence of the T$_1$-S$_0$ energy gap as a function of the \gls{CAS} size, while the \gls{CMO} approach has a more erratic behavior.
For example, although the \gls{CMO} approach appears to predict the T$_1$-S$_0$ energy gap exactly at a CAS(8,16) size, increasing the active space size to CAS(8,22) leads to a significant drop in the energy gap.
This suggests that simply increasing the active space size does not necessarily lead to a more accurate prediction of the energy gap using \glspl{CMO}.
\begin{table} 
  \centering
  \caption{Total energies (in \si{\hartree}) for \ce{O2} with CAS(16,32) using both \gls{CMO} and \gls{FNO} approaches as compared to \gls{CCSD-T} results.}
  \begin{tabular*}{\linewidth}{@{\extracolsep{\fill}} l S S S S @{}}
    \toprule
    State & \multicolumn{1}{c}{CMO} & \multicolumn{1}{c}{FNO} & \multicolumn{1}{c}{FNO + $\Delta E_{\text{FNO}}$} & \multicolumn{1}{c}{CCSD(T)} \\
    \midrule
    S$_0$ & \num{-149.775} & \num{-150.028} & \num{-150.122} & \num{-150.122} \\
    T$_1$ & \num{-149.824} & \num{-150.071} & \num{-150.167} & \num{-150.169} \\
    \cmidrule{2-5} 
    T$_1$ - S$_0$ & \num{-0.049} & \num{-0.043} & \num{-0.045} & \num{-0.047} \\
    \bottomrule
  \end{tabular*}
  \label{tab:o2-cmo-vs-fno}
\end{table}

\begin{figure*}
  \centering
  \includegraphics[width=0.75\textwidth,trim=4 4 4 5,clip]{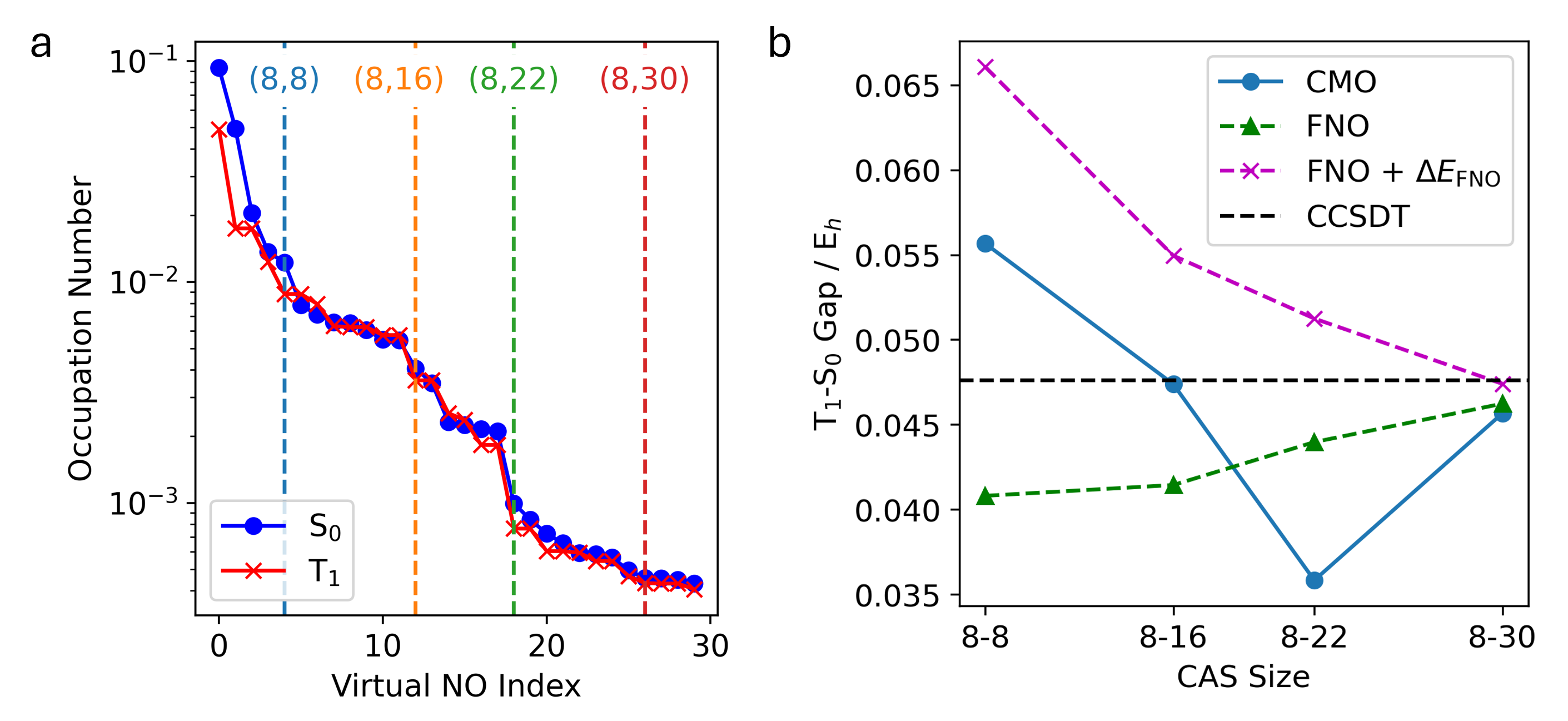}
  \caption{(a) \Glspl{NOON} for the virtual orbitals of \ce{O2} with the aug-cc-pVTZ basis set, for the S$_0$ and T$_1$ states.
    The orbitals used for \gls{CAS} sizes of (8,8), (8,16), (8,22), and (8,30) are shown by horizontal dashed lines, indicating that all orbitals to the left of that line are included in the given active space using the \gls{ZAPT-FNO} approach.
    (b) Convergence of the T$_1$-S$_0$ energy gap as a function of the active space size for \ce{O2} with the aug-cc-pVTZ basis set.}
  \label{fig:o2-gap-convergence}
\end{figure*}

The ZAPT-FNO approach achieves both systematic and accurate convergence, compared with the \gls{CMO} approach. The absolute energies from the \gls{iQCC} optimized results for \gls{ZAPT-FNO} converge to the \emph{true} ground state energy, while the \gls{CMO} approach converges to a value above the ground state.
Consider the CAS(16,32) active space size, which reproduces the T$_1$-S$_0$ energy gap to within \SI{2}{\milli\hartree} of the \gls{CCSD-T} value both for the \gls{CMO} and \gls{ZAPT-FNO} approaches, as shown in Table~\ref{tab:o2-cmo-vs-fno}.
We can show that the apparent accuracy of the \gls{CMO} T$_1$-S$_0$ gap is only due to favorable error cancellation, rather than a more accurate total energy of either the T$_1$ or S$_0$ state.
We can observe this error cancellation by considering the absolute total energies of the T$_1$ and S$_0$ states individually, as shown in Table~\ref{tab:o2-cmo-vs-fno}.
Specifically, the S$_0$ energy computed with the \gls{FNO}+$\Delta E_{\text{FNO}}$ correction matches the \gls{CCSD-T} reference value exactly, and the T$_1$ energy lies within \SI{2}{\milli\hartree} of the reported \gls{CCSD-T} reference.
In contrast, the \gls{CMO}-based calculation overestimates both the S$_0$ and T$_1$ energies by approximately \SI{347}{\milli\hartree} and \SI{345}{\milli\hartree}, respectively, relative to \gls{CCSD-T}.
This is a similar effect shown in Fig.~\ref{fig:o2-gap-convergence}b, where the \gls{CMO} approach appears to converge to the correct T$_1$-S$_0$ gap at CAS(8,16), but subsequently exhibits non-monotonic behavior as the active space size is increased.

These results show that the \gls{CMO} orbital selection is heavily dependent on \emph{which} orbitals are included in the active space, whereas the \gls{ZAPT-FNO} approach shows smooth convergence simply with the \emph{number} of orbitals in the CAS.
This provides more control over the quality of the T$_1$-S$_0$ gap in the \gls{ZAPT-FNO} case.
Furthermore, we have the capability to recover the frozen out energies using the $\Delta E_{\text{FNO}}$ correction, which leads to total energies in the \gls{ZAPT-FNO} case which are within chemical accuracy of \gls{CCSD-T}.
Thus, the \gls{ZAPT-FNO} approach offers a more systematic route to convergence of both the T$_1$–S$_0$ excitation gap and the absolute total energies of each state.

\subsection{\ce{CH2} bond dissociation case study: from weak to strong correlation regimes}
\label{sec:ch2-triplet-singlet-gap}

The dissociation of a single \ce{H} atom from \ce{CH2} is a particularly interesting case to test the \gls{ZAPT-FNO} approach, as the singlet and triplet states undergo a level crossing during bond dissociation that is only correctly described in classical simulations by multireference \gls{CI} and very large augmented basis sets~\cite{kalemos2004ch2}.
Recently, \citet{liepuoniute2025quantum} employed a quantum-centric approach to compute the T$_1$-S$_0$ energy gap across this dissociation path for \ce{CH2} using the \gls{SQD} method with \num{52} qubits.
Their results show that the \gls{SQD} method with a CAS(6,23) and the cc-pVDZ basis set is able to reproduce the T$_1$-S$_0$ energy gap within \SI{5}{\milli\hartree} of the experimental value of \SI{14.4}{\milli\hartree}~\cite{bunker1986potential}, but struggles to reproduce the gap at \ce{C-H} lengths between \SIrange[range-phrase ={ and }]{2.0}{2.5}{\angstrom}, where static correlation dominates the wavefunction.

This challenge of recovering static correlation, and utilizing a small active space with a larger basis set is exactly the type of problem that the \gls{ZAPT-FNO} approach is designed to address.
We compute the T$_1$-S$_0$ energy gap for \ce{CH2} using the \gls{ZAPT-FNO} approach with the aug-cc-pVTZ basis set, at CAS(6,23), and compare the results with the \gls{CASCI} energy gap computed using the same \gls{CAS}.
The experimental $^1$A$_1$ singlet state, and $^3$B$_1$ triplet state geometry of \ce{CH2} were used as starting points for all calculations of \ce{CH2}.
The singlet state geometry has a bond length of \SI{1.11}{\angstrom}, and \ce{H-C-H} bond angle of \SI{102.4}{\degree}, and the triplet state geometry has a bond length of \SI{1.09}{\angstrom}, and \ce{H-C-H} bond angle of \SI{135.3}{\degree}~\cite{liepuoniute2025quantum,harrison1974structure,bunker1986potential}.
All \ce{CH2} Hamiltonians at each bond length and spin state were optimzed with iQCC, using \num{10} iterations of \num{10} entanglers at a maximum order of \num{8}, and a final step with \num{200000} entanglers at a maximum order of \num{1}.

As shown in Fig.~\ref{fig:ch2-bond-length-vs-energy}a, the \gls{ZAPT-FNO} approach predicts the location of the level crossing at \SI{1.46}{\angstrom}, both with \gls{CASCI} and \gls{iQCC}.
The T$_1$-S$_0$ energy gap computed for a single H bond dissociation is within \SI{1}{\milli\hartree} of the \gls{CASCI} value across all bond lengths from \SIrange[range-phrase={ to }]{0.7}{3.0}{\angstrom}, and follows the same trend as the \gls{CASCI} energy gap, as shown in Fig.~\ref{fig:ch2-bond-length-vs-energy}a (right panel).
This pronounced improvement in the quality of the total energies comes at no extra cost in the total number of qubits, as we are able to increase the basis set quality while simultaneously maintaining a small \gls{CAS} size.
Unlike the cc-pVDZ \gls{CMO} approach, which predicts T$_1$-S$_0$ energy gaps which diverge up to \SI{30}{\milli\hartree} from the \gls{CASCI} value, the \gls{ZAPT-FNO} approach is able to recover the T$_1$-S$_0$ energy gap within \SI{1}{\milli\hartree} across the entire bond dissociation path, including in the strongly correlated regions at stretched bond lengths between \SIrange[range-phrase={ and }]{2.0}{3.0}{\angstrom}, where the cc-pVDZ \gls{CMO} approach used by~\citet{liepuoniute2025quantum} fails reproduce the same energy gap.

The only divergence from chemical accuracy occurs when incorporating the $\Delta E_{\text{FNO}}$ correction, to compute the T$_1$-S$_0$ energy gap at larger bond lengths \SIrange[range-phrase={ and }]{2.5}{3.0}{\angstrom}.
We can explain this discrepancy by considering the limitations of \gls{ZAPT2} perturbation theory.
Using the \ce{Li2} molecule as an example, in Appendix~\ref{sec:recommendation}, we find that at larger bond lengths beyond twice the equilibrium bond length ($2r_e$), the \gls{ZAPT2} perturbation theory overestimates the occupation number of virtual orbitals.
For \ce{CH2}, the equilibrium bond length in the singlet state is \SI{1.11}{\angstrom} so $2r_e = \SI{2.22}{\angstrom}$, which explains the difference between the $E_{\text{iQCC}}^{\text{FNO}}$ and $E_{\text{iQCC}}^{\text{FNO}}$+$\Delta E_{\text{FNO}}$ at larger bond lengths \SIrange[range-phrase={ and }]{2.5}{3.0}{\angstrom}.
Thus at extended bond lengths beyond 2$r_e$, the $\Delta E_{\text{FNO}}$ correction will no longer provide a reliable estimate of the correlation energy from the frozen virtual orbitals, and therefore should not be used to correct the final energy.
\begin{figure*} 
    \centering
    \includegraphics[width=0.75\textwidth,trim=4 4 4 10,clip]{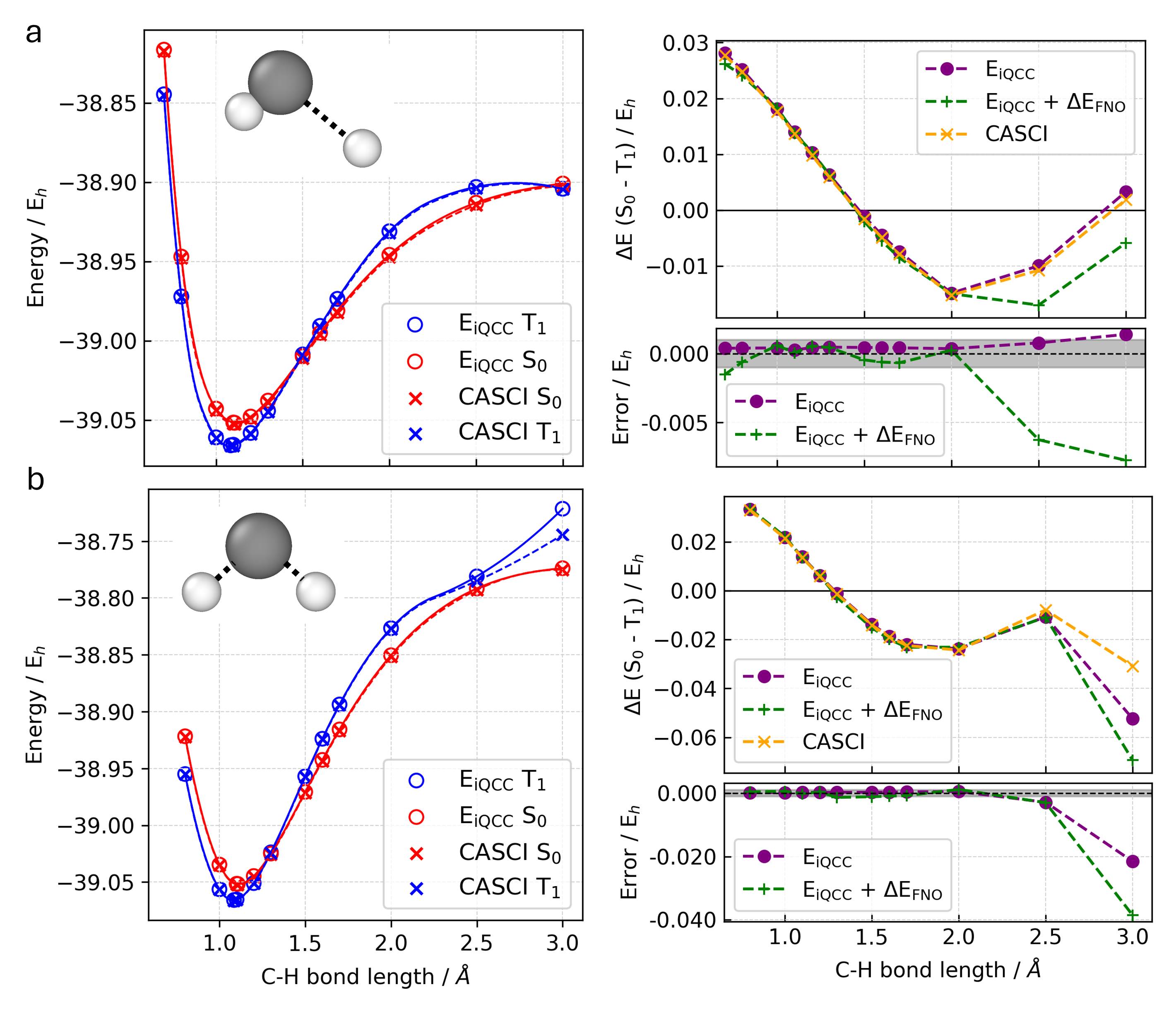}
    \caption{(a) Bond dissociation energy path of a breaking of a single \ce{C-H} bond in the \ce{CH2} molecule (left).
      T$_1$-S$_0$ energy gap across the dissociation pathway for breaking a single \ce{C-H} bond in \ce{CH2} (right).
      (b) Bond dissociation energy path of the symmetric stretching of both \ce{C-H} bonds in the \ce{CH2} molecule (left).
      T$_1$-S$_0$ energy gap across the dissociation pathway for symmetric stretching of both \ce{C-H} bonds in \ce{CH2} (right).
      The \gls{CASCI} calculations are performed on the \gls{ZAPT-FNO} selected orbitals also used to produce the qubit Hamiltonian for the \gls{iQCC} optimization.
      Chemical accuracy of \SI{1}{\milli\hartree} relative to the \gls{CASCI} calculations is indicated by the grey box in the right panel for both (a) and (b).
      Dashed lines are shown to guide the eye.}
    \label{fig:ch2-bond-length-vs-energy}
\end{figure*}

Given the success of the \gls{ZAPT-FNO} approach for a single bond dissociation, we now consider the more challenging problem of the symmetric stretching of both \ce{C-H} bonds in \ce{CH2}, which is shown in Fig.~\ref{fig:ch2-bond-length-vs-energy}b.
Once again keeping the active space size fixed at CAS(6,23), with the aug-cc-pVTZ basis set, we compute the T$_1$-S$_0$ energy gap across \ce{C-H} bond lengths from \SIrange[range-phrase={ to }]{0.8}{3.0}{\angstrom}.
Impressively, the \gls{ZAPT-FNO} approach is able to recover the T$_1$-S$_0$ energy gap within \SI{1}{\milli\hartree} of the \gls{CASCI} value across \ce{C-H} bond lengths up to \SI{2.5}{\angstrom}, (Fig.~\ref{fig:ch2-bond-length-vs-energy}b, right panel).
The location of the level crossing in the symmetric stretching of both \ce{C-H} bonds is at \SI{1.26}{\angstrom}, which is within \SI{0.15}{\angstrom}, of the spin-adiabatic transition state computed value of \SI{1.11}{\angstrom}~\cite{tao2020constructing}.

Beyond \SI{2.5}{\angstrom}, we enter the true dissociation regime ($>2r_e$), in which static correlation dominates the wavefunction, and the single reference nature of \gls{iQCC} is no longer able to capture the multireference character of the wave function.

As \gls{CASCI} is able to naturally treat multiple wavefunction configurations, we can conclude that the limitation here is with the nature of \gls{iQCC}, and indeed any single reference \gls{QCC} or eigensolver approach. The triplet-singlet gap computed between the experimental $^1$A$_1$~\cite{petek1989analysis} and $^3$B$_1$ states~\cite{bunker1986potential} using the \gls{ZAPT-FNO} approach is \SI{14.17}{\milli\hartree} ($E_{\text{iQCC}}^{\text{FNO}}$), which is within \SI{0.5}{\milli\hartree} of the experimental value of \SI{14.4}{\milli\hartree}~\cite{bunker1986potential}, as shown in Table~\ref{tab:ch2-energy-gaps}.
This is a significant improvement not only over the results by~\citet{liepuoniute2025quantum} using the \gls{SQD} method with the aug-cc-pVDZ basis set, but also over the results of the \gls{CMO} approach using the same basis set (aug-cc-pVTZ) at CAS(6,23), which predicts a T$_1$-S$_0$ energy gap of \SI{28.17}{\milli\hartree}.

We finally consider the quadruple zeta quality basis set, aug-cc-pVQZ, in order to test how well a larger basis set will describe the T$_1$-S$_0$ gap in \ce{CH2}.
We found that the \gls{iQCC} protocol used for the aug-cc-pVTZ basis with \num{10} steps of \num{10} entanglers set did not result in amplitudes below \num{0.015}, and therefore we optimized all Hamiltonians from the aug-cc-pVQZ basis set for \num{20} steps of \num{10} entanglers each at a maximum order of \num{8}, before the final step with \num{200000} entanglers.
As shown in Table~\ref{tab:ch2-energy-gaps}, we find that an active space size of CAS(6,23) is not sufficient to reproduce the experimental value of the energy gap, even at the \gls{CASCI} level of theory.
Notably, we find that the $E_{\text{iQCC}}^{\text{FNO}}$ energy gap agrees within \SI{0.5}{\milli\hartree} of the \gls{CASCI} value, however the $\Delta E_{\text{FNO}}$ correction, results in an energy gap value of \SI{7.690}{\milli\hartree}, which is \SI{7}{\milli\hartree} from the \gls{CASCI} value of \SI{13.142}{\milli\hartree}.
{This result highlights a particular point of importance about applying the $\Delta$E$_{\mathrm{FNO}}$ correction; in cases where the active space is aggressively truncated relative to the size of the basis set, it is possible for the overestimate the $\Delta$E$_{\mathrm{FNO}}$ correction.}
{This indicates that there is a large amount of dynamical correlation present in the frozen virtual orbitals.}
{In the case of aug-cc-pVQZ CAS(6,23), the perturbative correction to the singlet state is significantly overestimated compared to the triplet state, leading to an artificial narrowing of the energy gap.}

One of the major advantages of using \gls{iQCC} is that we can additionally test larger active spaces at very little additional computational cost~\cite{ryabinkin2023efficient}. 
Therefore, we chose to consider the T$_1$-S$_0$ energy gap for \ce{CH2} at a larger active space size of CAS(6,30) with the aug-cc-pVQZ basis set.
In this case, we find that the $E_{\text{iQCC}}^{\text{FNO}}$ energy gap is \SI{15.677}{\milli\hartree}, which is within \SI{1.5}{\milli\hartree} of the \gls{CASCI} value of \SI{15.412}{\milli\hartree}, and the $\Delta E_{\text{FNO}}$ correction leads to a final energy gap of \SI{14.534}{\milli\hartree}, which is within \SI{0.2}{\milli\hartree} of the experimental value of \SI{14.4}{\milli\hartree}~\cite{liepuoniute2025quantum}.

It is important to note that the greatest savings from the \gls{ZAPT-FNO} approach occur for basis sets with large, diffuse, or augmented orbitals such as the aug-cc-pVTZ and aug-cc-pVQZ basis sets.
We find that for the cc-pVDZ basis set, both the \gls{CMO} and \gls{ZAPT-FNO} approaches lead to similar T$_1$-S$_0$ energy gaps, as shown in Table~\ref{tab:ch2-energy-gaps}, which is to be expected.
This result is encouraging, as it implies that the Hamiltonian generated from \gls{ZAPT-FNO} approaches the same ground state energy as the equivalent size \gls{CMO} Hamiltonian, in basis sets without diffuse orbitals.
Thus, the advantage of the \gls{ZAPT-FNO} approach lies in the fact that it enables us to use larger basis sets, with similar \gls{CAS} sizes to achieve chemical accuracy.
The generality of \gls{ZAPT-FNO} to other quantum eigensolvers is further demonstrated in Appendix~\ref{sec:vqe-example}.

\begin{table*}
    \centering
    \caption{\ce{CH2} T$_1$-S$_0$ energy gap (in \si{\milli\hartree}) using the \gls{CMO} and \gls{ZAPT-FNO} approaches in different basis sets from the Dunning's family~\cite{dunning1989gaussian,kendall1992a,Peterson:1994/jcp/7410}.}
    \begin{tabular*}{\linewidth}{@{\extracolsep{\fill}} l l S S S S S @{}}
            \toprule
      Basis set & \gls{CAS} size & 
      \multicolumn{3}{c}{iQCC eigensolver} & 
      \multicolumn{1}{c}{CASCI eigensolver} & 
      \multicolumn{1}{c}{} \\ 
      \cmidrule(lr){3-5} \cmidrule(lr){6-6}
      & & 
      \multicolumn{1}{c}{CMO} & 
      \multicolumn{1}{c}{ZAPT-FNO} & 
      \multicolumn{1}{c}{\gls{ZAPT-FNO} + $\Delta E_{\text{FNO}}$} & 
      \multicolumn{1}{c}{\gls{ZAPT-FNO} + $\Delta E_{\text{FNO}}$} & 
      \multicolumn{1}{c}{Experiment\footnotemark[1]} \\
      \midrule
      cc-pVDZ     & (6,23) & 20.642 & 18.913 & 19.034 & 18.560 &  \\
      aug-cc-pVTZ & (6,23) & 28.170 & 14.166 & 13.897 & 13.880 &  \\
      aug-cc-pVQZ & (6,23) & 33.200 & 13.633 & 7.690  & 13.142 & \\
      aug-cc-pVQZ & (6,30) & 27.910 & 15.677 & 14.534 & 15.412 & 14.4 \\
      \hline
    \end{tabular*}
    \footnotetext[1]{Data from Ref.~\citenum{bunker1986potential}.}
    \label{tab:ch2-energy-gaps}
\end{table*}

\subsection{Efficient large-scale calculations with \gls{ZAPT-FNO} and \gls{iQCC}}
\label{sec:effic-large-scale}

\begin{table*}[!htb]
  \centering
  \caption{\ce{Ir(ppy)3} T$_1$-S$_0$ energy gap (in \si{\electronvolt}) using the \gls{CMO} and \gls{ZAPT-FNO} approaches.
    A double slash (//) separates an atomic basis set on an \ce{Ir} center from that on ligand atoms.}
  \begin{tabular*}{\linewidth}{@{\extracolsep{\fill}} l l S S S S @{}}
    \toprule
    Basis set & \gls{CAS} size & 
    \multicolumn{1}{c}{\gls{CMO}} &
    \multicolumn{1}{c}{\gls{ZAPT-FNO}} &
    \multicolumn{1}{c}{\gls{ZAPT-FNO} + $\Delta E_{\text{FNO}}$} &
    \multicolumn{1}{c}{Experiment\footnotemark[1]} \\
    \midrule
    LANL2TZ//6-31+G(d) & (40,40) & 2.897 & 2.638 & 2.412 &\multicolumn{1}{c}{{2.525}}  \\
    \bottomrule
  \end{tabular*}
  \footnotetext[1]{Data from Ref.~\citenum{sajoto2009temperature}.}
  \label{tab:irppy3-energy-gaps}
\end{table*}

\begin{figure*}[!htb]
  \centering \includegraphics[width=0.80\textwidth]{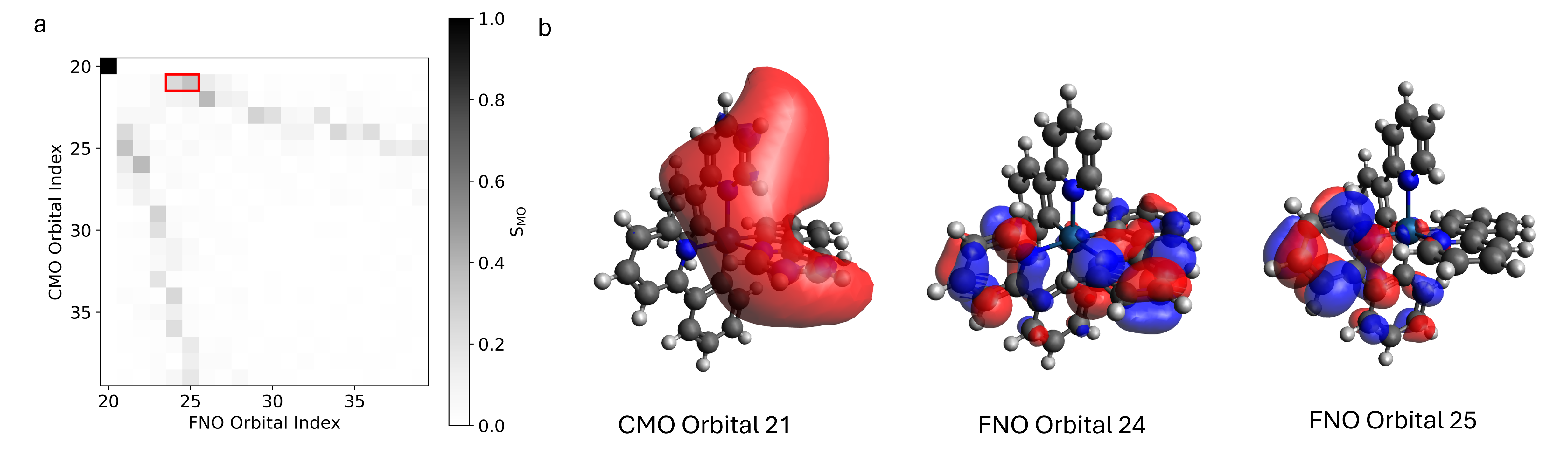}
  \caption{(a) Heatmap showing the orbital overlap matrix between the \gls{CMO} coefficients and semicanonicalized \gls{FNO} coefficients for the first \num{20} virtual orbitals of the \ce{Ir(ppy)3} complex (labelled orbitals \SIrange[range-phrase= --]{20}{39} as these are the last \num{20} orbitals in the CAS(40,40) set).
    This matrix captures how much each orbital from the \gls{FNO} set overlaps with the corresponding index in the \gls{CMO} set.
    The orbital overlap between CMO orbital 21 and FNO orbitals 24 and 25 which have the most overlap with CMO 21, is highlighted in a red box.
    (b)  Orbital \num{21} from the \gls{CMO} set which is a Rydberg-like orbital, compared with orbitals 24 and 25 from the \gls{FNO} set which are localized, and shows $s$ and $p$-like character on the carbon rings despite having the most orbital overlap spatially with CMO orbital 21.}
  \label{fig:irppy3-orbital-overlap}
\end{figure*}
The accurate prediction of triplet-singlet energy gaps is a fundamental requirement for the design and modelling of phosphorescent materials~\cite{genin2022estimating}.
We expect the \gls{ZAPT-FNO} approach to be well suited for optimizing the energy gaps in these materials, where large diffuse Rydberg orbitals dominate the virtual space.
To illustrate the potential of this approach, we consider the \ce{Ir(ppy)3} complex, which has an experimental T$_1$-S$_0$ gap of \SI{2.525}{\electronvolt}~\cite{sajoto2009temperature}.
The optimized T$_1$ geometry of \ce{Ir(ppy)3} was extracted from Ref.~\citenum{genin2022estimating}, and is provided as part of the Supplementary Information.

To demonstrate the capability of the method on a large phosphorescent material, we show that with \gls{ZAPT-FNO} we are able to use very large, augmented basis sets to correctly describe the electronic structure of Ir(ppy)$_3$, while maintaining a reasonable \gls{CAS} size.
Specifically, we compare results using both \gls{CMO} and \gls{ZAPT-FNO} orbital selection within the same LANL2TZ//6-31+G(d) basis set (see Table~\ref{tab:irppy3-energy-gaps}).
We use the LANL2TZ basis set and ECP on the Ir atom, and the 6-31+G(d) basis set on all other atoms, with a CAS(40,40) to compute the T$_1$-S$_0$ energy gap.
The use of this triple-zeta diffuse basis set is enabled by the \gls{ZAPT-FNO} approach's ability to efficiently truncate the virtual space while maintaining accuracy.
Without restricting the active space, the total number of atomic orbitals is \num{731} (\num{601} of which are in the virtual space), and thus our truncation to CAS(40,40) represents a reduction of \SI{97}{\percent} of the total virtual space.

The \num{80} qubit \ce{Ir(ppy)_3} Hamiltonian was optimized using iQCC with \num{20} steps of \num{15} entanglers each at a maximum order of \num{4}, followed by \num{20} steps with \num{200} entanglers each and a maximum order of \num{2}, and finally \num{750000} entanglers at a maximum order of \num{1}. The total energies extracted from iQCC are those computed with the second-order \gls{ENPT} correction~\cite{ryabinkin2021posteriori,epstein1926stark,nesbet1955configuration}, as previously recommended in Ref.~\citenum{genin2022estimating}.

As shown in Table~\ref{tab:irppy3-energy-gaps}, the \gls{ZAPT-FNO} + $\Delta E_{\text{FNO}}$ approach yields a T$_1$-S$_0$ gap of \SI{2.412}{\electronvolt}, which is within \SI{0.11}{\electronvolt} (\SI{4}{\milli\hartree}) of the experimental value.
In comparison, the \gls{CMO} approach with the same LANL2TZ//6-31+G(d) basis set and CAS(40,40) active space yields a gap of \SI{2.897}{\electronvolt}, deviating from experiment by \SI{0.37}{\electronvolt} (\SI{14}{\milli\hartree}).
This comparison demonstrates that the \gls{ZAPT-FNO} approach achieves a threefold reduction in error compared to \gls{CMO}, highlighting the superior orbital selection provided by the \gls{ZAPT-FNO} method for open-shell systems with large diffuse basis sets.

We can attribute some portion of the discrepancy between \gls{CMO} and \gls{FNO} gaps to the inability of the \gls{CMO} to capture the correlation energy in the virtual space, and the diffuse Rydberg character of the virtual orbitals.
To confirm this, we compute the overlap between the \glspl{CMO} and the \glspl{FNO} for the \ce{Ir(ppy)_3} complex.
Figure~\ref{fig:irppy3-orbital-overlap}a presents the computed overlap values for the first \num{20} virtual orbitals of the \ce{Ir(ppy)3} complex in the T$_1$ state.
Notably, the first orbital highlighted along the diagonal axis is shaded in dark black and therefore identical (or nearly identical) between the \gls{CMO} and \gls{FNO} approach.
However, beyond the first virtual orbital, the low magnitude of the matrix elements indicates the divergence between the two sets.
This lack of overlap suggests that the orbitals retained in the \gls{FNO} approach are qualitatively different from those from the \gls{CMO} virtuals, and we can visualize this difference in Fig.~\ref{fig:irppy3-orbital-overlap}b.
Figure~\ref{fig:irppy3-orbital-overlap}b compares CMO 21, which exhibits diffuse Rydberg-like character, with FNOs 24 and 25, which have the largest overlap with CMO 21 as shown in Figure~\ref{fig:irppy3-orbital-overlap}a.
While the overlap matrix indicates spatial overlap between CMO 21 and FNOs 24 and 25, the orbital character is fundamentally different: FNOs 24 and 25 show tightly localized $s$- and $p$-like character on the carbon rings of the ligands, rather than the diffuse, delocalized character of CMO 21.
Although the Rydberg orbital has a low energy, and is therefore included in the \gls{CMO} active space, it does not have a large contribution to the correlation energy, and therefore is excluded from the \gls{FNO} active space.
The \gls{ZAPT-FNO} approach instead prioritizes orbitals like FNOs 24 and 25, which capture correlation effects through their localized character rather than through diffuse contributions.
In fact, considering all \num{20} of the \gls{FNO} virtual orbitals we find that they are all highly localized, and none have diffuse or Rydberg character.
Thus, we can conclude that the \gls{ZAPT-FNO} approach is a tractable method to select high-quality active spaces in diffuse basis sets for large open shell systems by prioritizing orbitals based on their correlation energy contributions rather than orbital energy alone.

\section{Conclusions}
\label{sec:conclusions}

We have introduced the \Acrfull{ZAPT-FNO} strategy for orbital selection in quantum eigensolvers, and demonstrated its accuracy and efficiency for open-shell systems.
Across diverse test cases, including the \ce{O2} triplet-singlet gap, \ce{CH2} bond dissociation, and \num{260} electron \ce{Ir(ppy)3} complex, the \gls{ZAPT-FNO} approach consistently outperforms canonical molecular orbital (\gls{CMO}) selection, by achieving chemical accuracy with limited active space sizes.
For \ce{CH2}, we demonstrate that we can achieve chemical accuracy for the T$_1$-S$_0$ energy gap across a broad range of bond lengths, including in strongly correlated regimes.

The advantages of the \gls{ZAPT-FNO} approach are most pronounced for large, diffuse basis sets, such as aug-cc-pVTZ or LANL2TZ, where canonical virtual orbitals are dominated by diffuse orbitals that do not contribute significantly to the total correlation energy.
By enabling the use of these larger basis sets without expanding the active space, \gls{ZAPT-FNO} retains orbitals that have high natural occupation, while freezing out Rydberg orbitals.
Even for compact basis sets, the method performs comparably to \gls{CMO} selection, underscoring its robustness across regimes.
Limitations arise only at stretched bond lengths, where second-order perturbation theory is known to break down.

Overall, the extension of the \gls{FNO} approach to open-shell systems provides a robust framework for preparing compact, accurate active spaces for quantum simulations with complex electronic structures that require high-quality basis sets.
This makes \gls{ZAPT-FNO} a compelling choice for both classical and hybrid quantum–classical algorithms, such as \gls{iQCC}, which require compact active spaces, and opens the door to accurate quantum simulations of challenging open-shell and excited-state systems.

\section{Supplementary Material}

The structure of \ce{Ir(ppy)3} used in Section \ref{sec:effic-large-scale} is provided in an \texttt{.xyz} file format.

\section{Acknowledgments}

Co-author Dr. Ilya Ryabinkin passed away in December 2025 while this article was under review. His contributions to this work, and to the broader field of quantum and computational chemistry, have been significant and lasting. Dr. Harper would also like to acknowledge his excellent mentorship and guidance during this project. 

The authors would like to acknowledge Grant No. 14234 from the Next Generation Manufacturing Canada (NGen) Project and Innovative Solutions Canada (ISC) project  No. 202208-F0033-C00003 for funding this work.

\section{Data Availability}

The data that support the findings of this study are available from the corresponding author upon reasonable request. The ZAPT-FNO code is currently only available on our internal platform, please contact the authors for requests to access the code.

\FloatBarrier

\appendix

\section{\gls{ZAPT2} one-particle density matrix in the \gls{MO} basis}
\label{sec:zapt2-one-particle-density-matrix}

In order to compute the virtual-virtual part of the one-particle density matrix ($P^{(2)}_{ab}$), as required by Step 2 of the \gls{ZAPT-FNO} approach, we can use the same notation as in Section \ref{sec:methods}, and define the one-particle density matrix in the \gls{MO} basis as,
\begin{align}
  P^{(2)}_{ab} =
  & \sum_{i,j} \sum_{p}^{\text{s.v.}}  
    \frac{(ia|jp)[2(ib|jp) - (ip|jb)]}{D_{ij}^{ap} D_{ij}^{bp}} \notag \\
  &+ \sum_{c} \sum_{p,q}^{\text{d.s.}}
    \frac{(pa|qc)[C_{pq}(pb|qc) - (pc|qb)]}{D_{pq}^{ac} D_{pq}^{bc}} \notag \\
  &+ \sum_{x,y} \sum_{i}
    \frac{(ix|ya)(ix|yb)}{D_{iy}^{xa} D_{iy}^{xb}} \notag \\
  &+ \frac{1}{2} \sum_{i,x,y}
    \frac{(ix|xa)(iy|yb)}{D_i^a D_i^b}.
    \label{eq:1-particle-density-matrix}
\end{align}
where indices $ab$ only run over the virtual orbitals.
All other parts of the $P^{(2)}$ matrix can be computed as described in~\cite{fletcher2002gradient}, but are not necessary to compute for the purposes of the \gls{FNO} selection.
With Equations \ref{eq:energy-correction} and \ref{eq:1-particle-density-matrix} we can now compute the virtual orbitals for the \gls{ZAPT-FNO} approach, and the corresponding energy correction.

\section{Recommendation for using $\Delta E_{\text{FNO}}$ energy correction}
\label{sec:recommendation}

In the main text, in Fig.~\ref{fig:ch2-bond-length-vs-energy}, we consider the T$_1$-S$_0$ energy gap of \ce{CH2} at different bond lengths.
We find that the \gls{iQCC} energy with the $\Delta E_{\text{FNO}}$ energy correction leads to a more accurate prediction of the T$_1$-S$_0$ energy gap than the uncorrected \gls{iQCC} energy, only for bond lengths up to \SI{2.0}{\angstrom}.
In this section, we explore the cause of this divergence, and provide a recommendation for when the $\Delta E_{\text{FNO}}$ energy correction should lead to improved agreement in the T$_1$-S$_0$ energy gap, and when it is likely to diverge from the expected value, especially at stretched bond lengths.

To examine this in more detail, we consider the simpler case of stretched \ce{Li2}, in the cc-pVDZ basis set.
The $\Delta E_{\text{FNO}}$ energy correction is computed as the difference in energy between the two subsystems of the orbital space, those orbitals which are strongly correlated (in the active space) and those orbitals which are more weakly correlated (frozen).
However, at stretched geometries, the contribution of these more weakly correlated orbitals in the \gls{ZAPT2} approximation is overestimated, and therefore the $\Delta E_{\text{FNO}}$ energy correction no longer is able to accurately capture the correlation energy in the frozen virtual space.
We can see this more clearly, by examining the \glspl{NOON} of the virtual orbitals in the \ce{Li2} molecule, beyond the first strongly correlated virtual orbital.
In Fig.~\ref{fig:li2-virtual-orbital-occupations}, we show the \gls{NOON} of 10 virtual orbitals in the \ce{Li2} molecule at bond lengths between  \SIrange[range-phrase={ to }]{1.3}{8.0}{\angstrom} computed using both the \gls{ZAPT2} approach and using the \glspl{CMO} from \gls{CASCI} calculation.
From \SIrange[range-phrase={ to }]{1.3}{4.0}{\angstrom}, we observe similar trends in the \glspl{NOON} of both the \gls{ZAPT2} and \gls{CMO} approaches, but beyond \SI{4.0}{\angstrom}, the \gls{ZAPT2} \glspl{NOON} diverge from the \gls{CMO} \glspl{NOON}, and the occupation numbers of the virtual orbitals increase relative to the \gls{CMO} \glspl{NOON}.
As a result, the $\Delta E_{\text{FNO}}$ energy correction will diverge from the expected value, since the orbitals in the frozen virtual space are overrepresented in the calculation of $\Delta E_{\text{FNO}}$.
\begin{figure}[!h]
    \centering
    \includegraphics[width=0.49\textwidth]{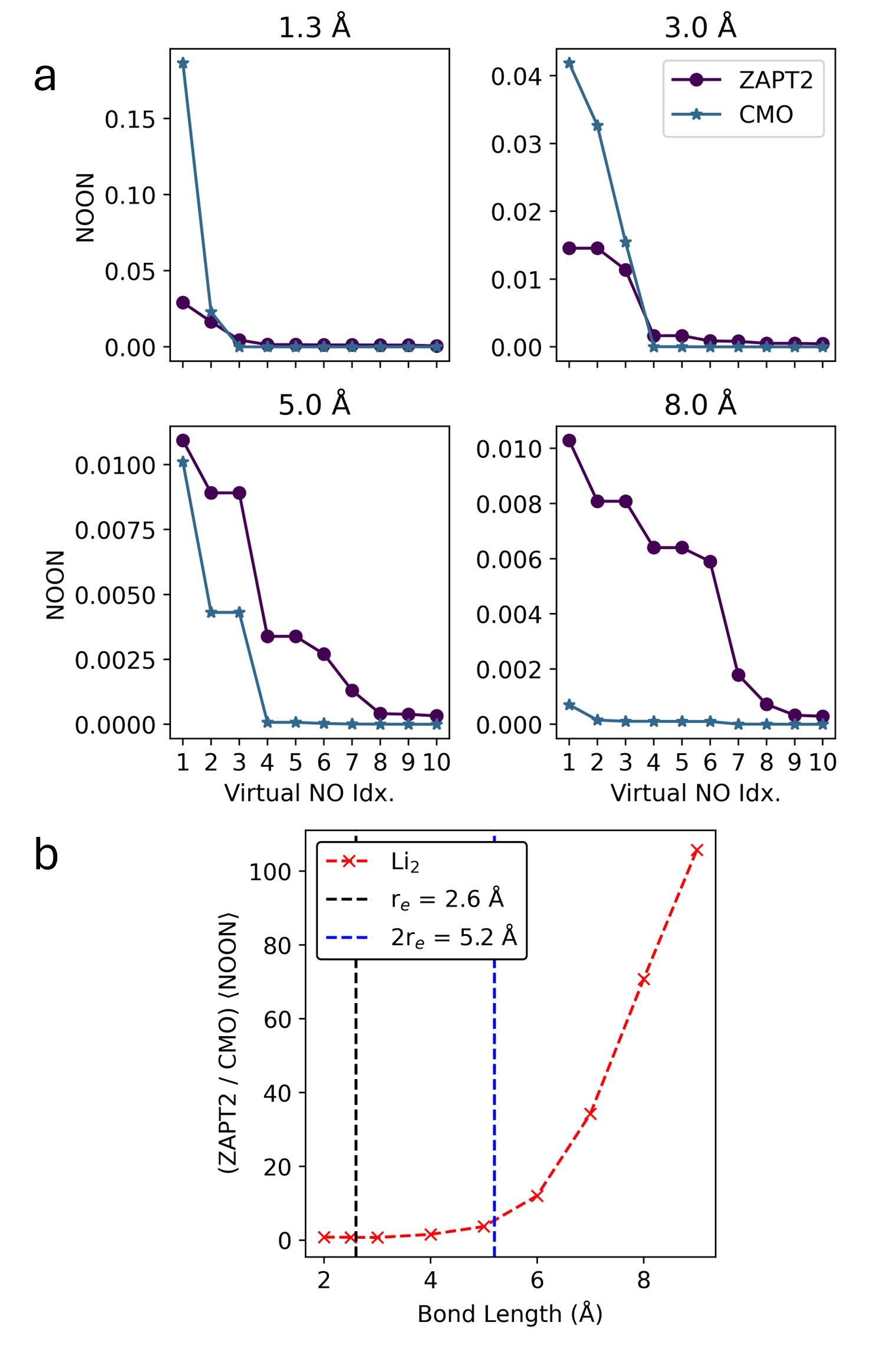}
    \caption{(a) Each tile shows the \glspl{NOON} of the first 10 virtual orbitals in the \ce{Li2} molecule, computed from the \gls{ZAPT2} P$^{(2)}$ density matrix  and the \gls{CASCI} computed \glspl{NOON} from the \gls{CMO} coefficients. 
      (b) The ratio of the \glspl{NOON} of all high lying virtual orbitals in the \ce{Li2} molecule at increasing \ce{Li-Li} distances.
      The equilibrium bond length (r$_e$) is \SI{2.6}{\angstrom}, and both $r_e$ and $2r_e$ are shown as vertical dashed lines.
      The ratio is computed as the average of all virtual \glspl{NOON} in the \gls{ZAPT2} approach divided by the \gls{NOON} of the virtual orbital in the \gls{CMO} approach.
      The ratio diverges at bond lengths beyond $2r_e$, indicating that the \gls{ZAPT2} approach overestimates the occupation number of the virtual orbitals in this regime.}
    \label{fig:li2-virtual-orbital-occupations}
\end{figure}

If we consider the ratio of the average of the virtual \glspl{NOON} across this bond stretching range, the divergence is more pronounced, as shown in Fig.~\ref{fig:li2-virtual-orbital-occupations}b.
At bond stretching lengths more than twice the equilibrium bond length ($2r_e = $~\SI{5.2}{\angstrom}), the ratio of the \glspl{NOON} diverges, as the virtual orbital occupations are overestimated by the \gls{ZAPT2} approach.
From \SIrange[range-phrase={ to }]{1.3}{4.0}{\angstrom}, we observe similar trends in the \glspl{NOON} of both the \gls{ZAPT2} and \gls{CMO} approaches, but beyond \SI{4.0}{\angstrom}, the \gls{ZAPT2} \glspl{NOON} diverge from the \gls{CMO} \glspl{NOON}, and the occupation numbers of the virtual orbitals increase relative to the \gls{CMO} \glspl{NOON}.
As a result, the $\Delta E_{\text{FNO}}$ energy correction will diverge from the expected value, since the orbitals in the frozen virtual space are overrepresented in the calculation of $\Delta E_{\text{FNO}}$.

\section{Computational Cost of ZAPT-FNO Workflow}
\label{sec:cpu-scaling}
Although the primary focus of the \gls{ZAPT-FNO} approach is to improve the accuracy of open-shell quantum computations, we have parallelized the \gls{ZAPT-FNO} code in order to be able to compute the virtual-virtual part of the one-particle density matrix ($P^{(2)}_{ab}$), as described in Appendix \ref{sec:zapt2-one-particle-density-matrix} more efficiently. 
As shown in Figure \ref{fig:cpu-vs-runtime}, we achieve reasonable speedup from serial to 16 CPU parallelization across the entire \gls{ZAPT-FNO} workflow on a single node. 
During the run, \texttt{pyscf} ROHF optimization is internally parallelized, and we have used the \texttt{mpi4py} approach to parallelize both the computation of the virtual-virtual part of the P$^{(2)}_{ab}$ density matrix and the $E^{(2)}_{\text{MP}}$ energy correction.

For all of the small molecules tested in this work, the runtimes and memory requirements, parallelized across 16 cores on a compute node featuring two Intel Xeon E5-2650 v3 processors (20 physical cores total), are shown in Table \ref{tab:runtimes}, averaged across 10 \gls{ZAPT-FNO} workflow computations.
The large 61 atom \ce{Ir(ppy)3} molecule, with the LANL2TZ//6-31+G(d) basis set required a large amount of RAM (\SI{1.02}{\tebi\byte}) in order to hold the 2-electron integrals in memory and therefore was run across 128 cores on a compute node featuring two AMD EPYC 7702 processors (128 physical cores total) with \SI{4}{\tebi\byte} of RAM.
Of course, this RAM bottleneck at large electron and orbital counts should be addressed in future implementations of the \gls{ZAPT-FNO} code.
However, as code optimization was not a major factor in our approach, and we are able to successfully run the T$_1$ and S$_0$ states in less than a week of wallclock time (see Table \ref{tab:runtimes}) we believe there is reasonable applicability across other similar sized systems.

\begin{table}[ht]
\centering
\caption{Timing and memory requirements for the ZAPT-FNO workflow for all molecules included in this work. For all molecules, except for \ce{Ir(ppy)3}, a set of 10 calculations were performed on 16 cores of a compute node with two Intel Xeon CPU E5-2650 v3 processors. The mean and standard deviation of the runtime and maximum memory usage was collected and is reported below. \ce{Ir(ppy)3} molecule required \SI{1.02}{\tebi\byte} of RAM in order to successfully construct the ZAPT-FNO orbitals, and therefore only a single run was performed on an AMD EPYC 7702 processor, parallelized over 128 cores.}
\begin{tabular}{l
                l
                S[table-format=6.2]
                S[table-format=6.2]
                S[table-format=6.2]}
\toprule
\textbf{Molecule} & \textbf{CAS size} & \textbf{Runtime (s)} & \textbf{Max Memory (GB)} \\
\midrule
Li$_2$     & (6,12)  & {\num{7.909} \,$\pm$\,\num{0.207}} & {\num{0.928} \,$\pm$\,\num{0.0042}} \\
CH$_2$ $^3$B$_1$     & (6,23)  & {\num{82.93} \,$\pm$\,\num{1.364}} & {\num{2.006} \,$\pm$\,\num{0.0158}} \\
CH$_2$ $^1$A$_1$     & (6,23)  & {\num{72.33} \,$\pm$\,\num{0.690}} & {\num{2.010} \,$\pm$\,\num{0.0082}} \\
O$_2$ T$_1$ & (8,22)  & {\num{126.99} \,$\pm$\,\num{2.860}} & {\num{2.177} \,$\pm$\,\num{0.0125}} \\
O$_2$ S$_0$ & (8,22)  & {\num{82.406} \,$\pm$\,\num{1.308}} & {\num{2.170} \,$\pm$\,\num{0.0176}} \\
H$_2$O$_2$ T$_1$ & (18,20) &{\num{425.17} \,$\pm$\,\num{5.860}} & {\num{4.495}\,$\pm$\,\num{0.5112}} \\
\hline
\ce{Ir(ppy)3} S$_0$ & (40,40)  & {\num{463417}} & {\num{761}} \\
\ce{Ir(ppy)3} T$_1$ & (40,40)  & {\num{626028}} & {\num{1020}} \\
\bottomrule
\end{tabular}
\label{tab:runtimes}
\end{table}

\begin{figure}[!h]
    \centering
    \includegraphics[width=0.49\textwidth]{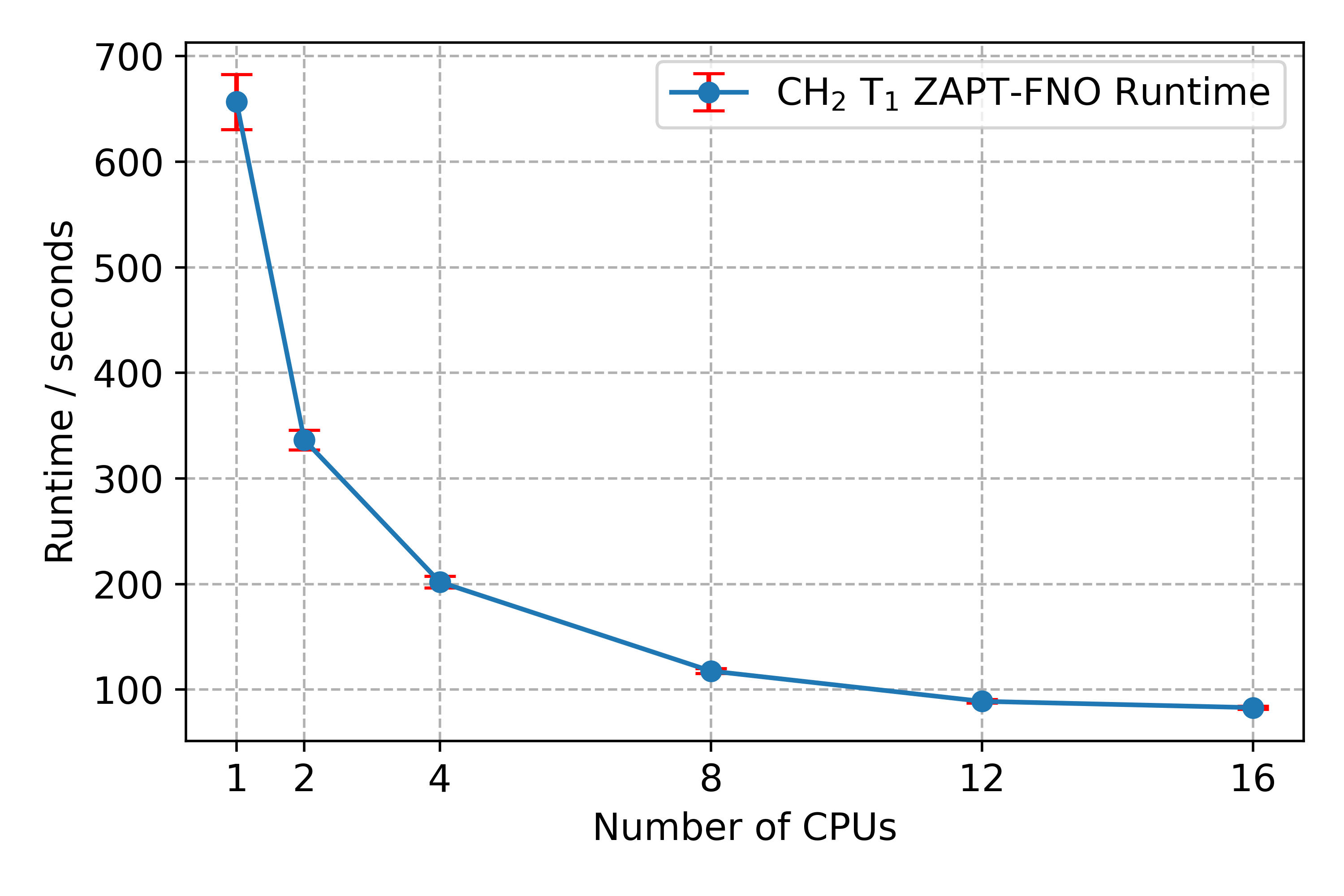}
    \caption{CPU scaling of the ZAPT-FNO approach from 1 to 16 CPUs for the T$_1$ state of CH$_2$. The runtime is calculated as the average over 10 runs, with the red error bars denoting the standard deviation across those runs. All calculations were carried out the Intel Xeon CPU E5-2650 v3 node. The runtime includes the entire ZAPT-FNO workflow from ROHF calculation to orbital construction to Hamiltonian generation including the \gls{JW} transformation \cite{nielsen2005fermionic}.}
    \label{fig:cpu-vs-runtime}
\end{figure}

\FloatBarrier

\section{ZAPT-FNO orbital selection with VQE}
\label{sec:vqe-example}

To demonstrate the generality of \gls{ZAPT-FNO} beyond \gls{iQCC} and \gls{CASCI} eigensolvers, we performed calculations using \gls{VQE} as implemented in Qiskit~\cite{qiskit2024},
using the \gls{UCCSD} ansatz with the parity mapping and the Sequential Least SQuares Programming optimizer (SLSQP)  \cite{kraft1988software} with a maximum of 200 iterations as implemented in \texttt{qiskit-nature}~\cite{qiskit-nature}. 
The \glspl{CMO} were obtained from PySCF~\cite{sun2020recent} ROHF/aug-cc-pVTZ calculations for the singlet ($^1$A$_1$) and triplet ($^3$B$_1$) states of CH$_2$, using the same geometries as described in Section \ref{sec:ch2-triplet-singlet-gap} of the main text, and the \glspl{FNO} were obtained using the \gls{ZAPT-FNO} workflow.

Due to the exponential scaling of resources with the number of qubits in the serial \texttt{qiskit} version, we were limited to testing CAS(6,6) and CAS(6,8) active spaces. 
Larger active spaces, such as the CAS(6,23) tested with \gls{iQCC} would be computationally prohibitive for the current \texttt{qiskit} infrastructure.

Table~\ref{tab:vqe_comparison} presents the VQE-computed energies and T$_1$-S$_0$ gaps for CH$_2$ using CAS(6,6) and CAS(6,8) active spaces. 
The results demonstrate that both the \gls{ZAPT-FNO} and \gls{ZAPT-FNO}+$\Delta E_\text{FNO}$ orbital selection strategies provide substantially more accurate T$_1$-S$_0$ gaps relative to experiment, as compared to \glspl{CMO} within the same active space. 
For CAS(6,8), the  \gls{ZAPT-FNO}+$\Delta E_\text{FNO}$ approach yields a gap of $13.46$ mE$_h$, compared to $30.96$ mE$_h$ for CMOs; the \gls{ZAPT-FNO}+$\Delta E_{\text{FNO}}$ result is within 0.94 mE$_h$ of the experimental value. 
The trends observed with VQE mirror those seen with classical CASCI and iQCC eigensolvers presented in Section \ref{sec:ch2-triplet-singlet-gap} of the main text, validating the generality of the ZAPT-FNO orbital selection strategy across different eigensolvers. 
Notably, these results demonstrate that \gls{ZAPT-FNO} provides a realistic path towards obtaining accurate T$_1$-S$_0$ gaps in significantly reduced active space sizes. 
Given the accuracy achieved within such a small active space, this also underscores the use of ZAPT-FNO orbital selection in the context of quantum eigensolvers where qubit counts are a bottleneck to chemical accuracy.

\renewcommand{\arraystretch}{1.8} 
\begin{table}[htbp]
\centering
\caption{VQE energies (E$_h$) and T$_1$-S$_0$ gap (mE$_h$) for CH$_2$ using CAS(6,6) and CAS(6,8) active spaces with the aug-cc-pVTZ basis set.}
\label{tab:vqe_comparison}
\begin{tabular}{llcccc}
\hline
\hline
 & & \multicolumn{3}{c}{VQE eigensolver} & \\
\cmidrule(lr){3-5}
CAS Size& State & CMO & ZAPT-FNO & \shortstack{\rule{0pt}{3ex}ZAPT-FNO \\ + $\Delta E_\text{FNO}$} & Expt. \\
\hline
\multirow{3}{*}{(6,6)} & S$_0$ & $-38.8960$ & $-38.9467$ & $-39.0756$ & \\
         & T$_1$ & $-38.9321$ & $-38.9682$ & $-39.0911$ & \\
         & T$_1$-S$_0$ & $36.17$ & $21.49$ & $15.54$ & 14.4 \\
\hline
\multirow{3}{*}{(6,8)} & S$_0$ & $-38.9016$ & $-38.9793$ & $-39.0796$ & \\
         & T$_1$ & $-38.9326$ & $-38.9889$ & $-39.0931$ & \\
         & T$_1$-S$_0$ & $30.96$ & $9.52$ & $13.46$ & 14.4 \\
\hline
\hline
\end{tabular}
\end{table}

\clearpage
\section*{References}
\label{sec:references}

\bibliographystyle{apsrev4-2}
\bibliography{apssamp}

\end{document}